\documentclass[12pt]{article}

\usepackage{amssymb,amsfonts,amsmath}
\usepackage{graphicx} 
\usepackage{indentfirst}
\usepackage{epstopdf}

\parskip=\medskipamount

\newcommand {\cD}{{\cal D}}

\newcommand {\cF}{{\cal F}}
\newcommand {\cG}{{\cal G}}

\newcommand {\cL}{{\cal L}}
\newcommand {\cM}{{\cal M}}
\newcommand {\cN}{{\cal N}}

\newcommand {\cP}{{\cal P}}

\newcommand {\cR}{{\cal R}}

\newcommand {\cV}{{\cal V}}
\newcommand {\cW}{{\cal W}}

\newcommand{\bA}{{\bf A}}

\newcommand{\bD}{{\bf D}}

\newcommand{\bL}{{\bf L}}

\def\a{\alpha}
\def \bi{\bibitem}

\def\b{\beta}

\def\d{\delta}
\def\e{\epsilon}
\def\f{\phi}
\def\g{\gamma}
\def\G{\Gamma}

\def\j{\psi}
\def\k{\kappa}
\def\l{\lambda}
\def\m{\mu}

\def\p{\pi}
\def\q{\theta}

\def\s{\sigma}
\def\t{\tau}

\def\x{\xi}
\def\z{\zeta}
\def\D{\Delta}
\def\F{\Phi}
\def\J{\Psi}

\def\O{\Omega}

\def\S{\Sigma}
\def\U{\Upsilon}
\def\X{\Xi}

\newcommand{\bd}{{\dot{\beta}}}                            
\newcommand{\ve}{\varepsilon}                            

\newcommand{\pa}{\partial}                           
\newcommand{\hf}{\frac12}

\newcommand{\vf}{\varphi}
\newcommand{\sect}[1]{\setcounter{equation}{0}\section{#1}}

\newcommand{\be}{\begin{equation}}
\newcommand{\ee}{\end{equation}}
\newcommand{\bea}{\begin{eqnarray}}
\newcommand{\eea}{\end{eqnarray}}
\newcommand{\non}{\nonumber}
\newcommand{\1}{\underline{1}}
\newcommand{\2}{\underline{2}}

\def\dt#1{{\buildrel {\hbox{\LARGE .}} \over {#1}}}    

\newcommand{\bm}[1]{\mbox{\boldmath$#1$}}

\def\double #1{#1{\hbox{\kern-2pt $#1$}}}

\begin{document}

\begin{titlepage}

\begin{flushright}
UMD-PP-05-48\\
hep-th/0507176\\
July, 2005\\ 
Revised version: January, 2006
\end{flushright}
\vspace{5mm}

\begin{center}
{\LARGE \bf  
On Five-dimensional  Superspaces}
\end{center}

\begin{center}

{\large  
Sergei M. Kuzenko\footnote{{kuzenko@cyllene.uwa.edu.au}}${}^\ddagger$
and
William D. Linch, 
III\footnote{{ldw@physics.umd.edu}}${}^,$\footnote{Address after September 1, 2005:
{\rm C. N. Yang Institute for Theoretical Physics}
{\rm  and the Department of Mathematics,}
{\rm State University of New York, Stony Brook, NY 11794.}
}
${}^\dagger$} \\
\vspace{5mm}

${}^\ddagger$\footnotesize{
{\it School of Physics M013, The University of Western Australia,\\
35 Stirling Highway, Crawley W.A. 6009, Australia}}  \\
~\\
${}^\dag$\footnotesize{
{\it Center for String and Particle Theory\\
Department of Physics, University of Maryland\\
College Park, MD 20742-4111 USA}}\\
~\\

\vspace{2mm}

\end{center}
\vspace{5mm}

\begin{abstract}
\baselineskip=14pt
\noindent
Recent  one-loop calculations of certain supergravity-mediated
quantum corrections in supersymmetric brane-world models 
employ either the component formulation (hep-th/0305184) or  
the superfield formalism with only half of  the bulk supersymmetry 
manifestly realized (hep-th/0305169 and hep-th/0411216).  
There are reasons to expect, however,  that 5D supergraphs 
provide a more efficient setup to deal with these 
and more involved (in particular, higher-loop) calculations. 
As a first step toward elaborating such supergraph techniques,  
we develop in this letter a manifestly supersymmetric formulation 
for 5D globally supersymmetric theories with eight supercharges. 
Simple rules are given to reduce 5D superspace actions to 
a hybrid form which keeps manifest only the 4D, $\cN=1$ 
Poincar\'e  supersymmetry. (Previously, such hybrid actions 
were carefully  worked out by rewriting the component 
actions in terms of simple superfields).  
To demonstrate the power of this formalism 
for model building  applications, two families of 
off-shell  supersymmetric nonlinear sigma-models 
in five dimensions are presented  (including those with 
cotangent bundles of K\"ahler  manifolds as target spaces). 
We elaborate, trying to make our presentation 
maximally  clear and self-contained,  
on the techniques of 
5D harmonic and projective  superspaces used at some stages 
in this letter.  
 \end{abstract}

\vfill
\end{titlepage}

\newpage
\setcounter{page}{1}
\renewcommand{\thefootnote}{\arabic{footnote}}
\setcounter{footnote}{0}

\begin{figure}[h] 
   \centering
\begin{center}
   \includegraphics[width=2.5in]{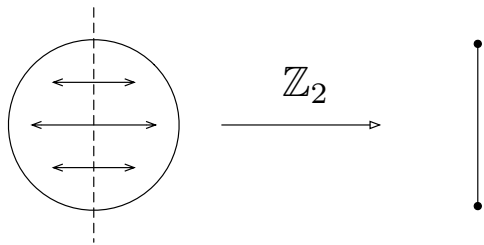} 
   \caption{The orbifold construction.}
   \label{fig:Orbifold}
\end{center}
\end{figure}

\begin{figure}[h] 
   \centering
   \includegraphics[width=5in]{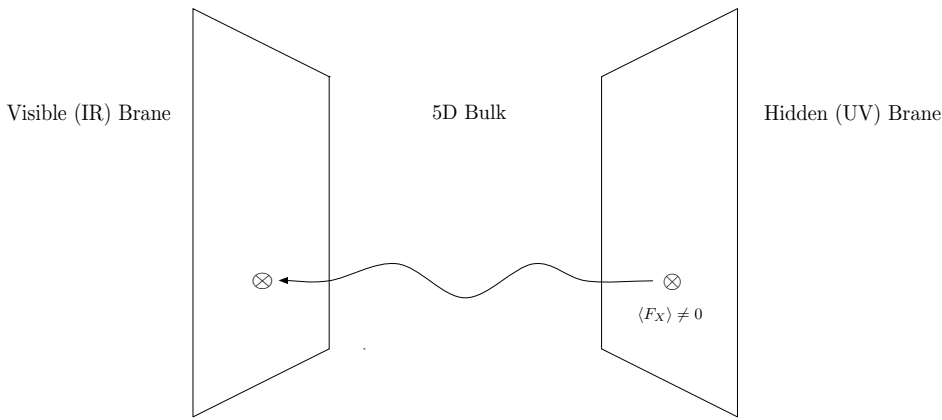} 
   \caption{The 5D brane-world scenario of the gravitational mediation 
of supersymmetry breaking in the standard model. The $F$-term of 
a chiral field $X$ on the hidden brane acquires a vacuum excitation
value. This is then communicated to the standard model (visible) brane 
by gravitational messengers propagating through the bulk.}
   \label{fig:Branes}
\end{figure}

\sect{Introduction}
Supersymmetric field theories in dimensions higher than 
the four accessible in our everyday experiences have been 
contemplated for many years now. Besides being forced on us by 
our current  understanding of superstring theory, it has 
also proven to be of possible phenomenological importance in 
which discipline such theories are known as supersymmetric 
``brane-world'' models \cite{Ant,HW,DvSh,RanSun}. 
Currently, this type of theory is being used 
by various groups in attempts to implement supersymmetry breaking 
in a manner consistent with the stringent bounds coming from flavor
changing neutral currents. One particular application uses a 
supersymmetric gravitational theory in a five-dimensional spacetime 
of which the ``extra'' spacial dimension is a compact interval of some
length\footnote{An elegant way of constructing such a space is to start 
with a 5D Minkowski space 
and toroidally compactify   
one of the spacial directions, 
producing a space of topology
${\mathbb R}^{4}\times {S}^1$. 
One subsequently defines a non-free
action of ${\mathbb Z}_2$ on the circle which has two antipodal fixed 
points (i.e. a reflection through a ``diameter'')
and mods the circle by this action. Since the action was not free, 
the quotient $S^1/{\mathbb Z}_2$ is not a smooth manifold 
(figure \ref{fig:Orbifold}). Nevertheless it is a manifold-with-boundary 
diffeomorphic to the closed finite interval $[0,1]$. 
The ``orbifold'' ${\mathbb R}^{4}\times ({S}^1/{\mathbb Z}_2)$ has 
as boundary two hyperplanes (the ``orbifold fixed planes''), each isometric 
to a 4D  Minkowski space,
at the fixed points 
of the ${\mathbb Z}_2$ action. 
} 
$\ell$ \cite{RanSun,multiple}. 
At each end of the interval is a copy of 4D Minkowski spacetime 
(the branes). Brane world models consist of postulating that the 
standard model of particle physics is localized on one of these 
branes (the ``visible'' or ``infra-red'' brane) while other fields propagate 
in the interior (``bulk'') of the 5D spacetime or on the other ``hidden'' 
(or ``ultraviolet'') brane (see figure \ref{fig:Branes}).
In the particular model under consideration, supersymmetry 
is broken on the hidden brane by allowing the $F$- (or $D$-term
 \cite{Gregoire:2005jr}) of some field localized on that brane 
to acquire a vacuum expectation value. This is then communicated 
to our brane by super-graviton loops in the bulk which induce the 
breakdown of supersymmetry on the standard model brane. The final
result is a four-dimensional effective action for the standard model 
fields on the visible brane.

Needless to say, calculations such as that of 
gravitational loop corrections are difficult 
to perform in components (although it was done  in 
\cite{Rattazzi:2003rj} at one loop). 
On the other hand, it is commonly believed
that the development of a 
full-fledged 5D superspace formulation has the annoying drawback 
that the result, which is desired to be a four-dimensional effective action,
is given in a complicated form. More exactly, the output of such an effort 
is manifestly supersymmetric in five dimensions and must be dimensionally 
reduced in the final stages of the calculation.  
${}$For  this reason a ``hybrid'' formalism was developed for 
supergravity in five dimensions which keeps manifest only 
4D, $\cN=1$  super-Poincar\'e invariance 
\cite{Linch:2002wg}. This hybrid is given in terms of supergravity 
prepotentials which allows one to apply the powerful supergraph 
techniques necessary 
for perturbative quantum calculations. Indeed, in \cite{Buchbinder:2003qu} 
it was used to compute, in a more econmical way  
than in the component approach of  \cite{Rattazzi:2003rj}, 
the leading gravity loop contribution to supersymmetry
breakdown described above in a very simple way. 

Although the formalism was successfully extended to allow 
a ``warping'' of the extra dimension and the gravity-mediation 
scenario investigated in this background \cite{Gregoire:2004nn}, 
it has a major drawback which arises as follows. This 
approach is essentially a superfield Noether procedure 
in which  one  starts with a linearized 
supergravity action, and then tries to reconstruct 
interaction terms,  
order by order, by consistently deforming 
the gauge transformations, etc.
Usually the Noether procedure can  be completed 
if it requires a finite number of iterations, as  
is the case with polynomial actions. 
But superfield supergravity is a highly nonlinear theory 
in terms of its prepotentials 
(see \cite{GGRS,BK} for reviews).
As a result, the limitations of this hybrid approach are called into question.
More importantly, 
it turns out to be difficult to discover the rules 
governing the coupling of this theory to other matter fields in the bulk. 
In the end, we are forced to turn to the known (full-fledged)
off-shell formulations for 5D simple supergravity,\footnote{We 
prefer to use 
the term ``5D simple  supergravity,'' since   
in the literature 5D simple supersymmetry is  
called sometimes  $\cN=1$ and sometimes $\cN=2$, 
depending upon taste and background.} 
with or 
without supersymmetric matter, 
in the hope of deducing
a
useful superfield formulation.

Off-shell 5D simple supergravity was  sketched
in superspace, a quarter  century ago,
by Breitenlohner and Kabelschacht \cite{Breitenlohner}  
and  Howe  \cite{Howe} 
(building on a related work \cite{HL}).
More recently, it was carefully  elaborated 
by Zucker \cite{Zucker}  at the component level, and finally perfected 
in \cite{Ohashi,Bergshoeff} within the superconformal tensor calculus.

Using the results of the 5D superconformal tensor calculus
for supergravity-matter systems, 
one can develope a hybrid $\cN=1$ superspace formalism 
by fitting the component multiplets into superfields. 
Such a program has been carried out in \cite{Correia:2005sr}. 
Although useful for tree-level phenomenological applications, 
we believe this approach is not the optimum (economical) 
formulation for doing supergraph loop calculations.
The point is that the superconformal tensor calculus 
usually corresponds to a Wess-Zumino gauge in superfield 
supergravity. But  such gauge conditions are impractical 
as far as supergraph calculations are concerned.

When comparing the superconformal tensor calculus
for 5D simple supergravity
\cite{Ohashi,Bergshoeff} 
with that for 4D, $\cN=2$ 
and 6D, ${\cal N}=(1,0)$
supergravities
(see \cite{Ohashi,Bergshoeff} for 
the  relevant references), it is simply  staggering how similar 
these formulations are, modulo some fine details.
${}$From the point of view of a superspace practitioner, 
the reason for this similarity is that the three versions
of superconformal calculus are generated from 
(correspond to a Wess-Zumino gauge for)
a harmonic superspace formulation for 
the corresponding supergravity theory, 
and such  harmonic superspace formulations\footnote{The
harmonic superspace formulation
 for 4D, $\cN=2$ supergravity is reviewed in 
the book \cite{GIOS}. For the case of 6D, 
$\cN=(1,0)$ 
supergravity, such a formulation was constructed 
in \cite{Sokatchev},  and it can be used 
to derive a relevant formulation for 5D simple supergravity 
by dimensional reduction.}
look almost identical in the space-time dimensions
4, 5 and 6, again modulo fine details. 
${}$For example, 
independent of the space-time dimension, 
the Yang-Mills supermultiplet is described by 
(formally) the same gauge superfield $\cV^{++}$, 
with the same gauge freedom $\d \cV^{++} = -\cD^{++} \lambda$, 
and with the  same Wess-Zumino-type  gauge
$$
{\rm i}\,\cV^{++}= 
\q^+ \G^{m} \q^+ A_{m}({x})
+\q^+ \G^{5} \q^+ A_{5}({x})
+\q^+ \G^{6} \q^+ A_{6}({x}) ~+~O(\q^3)~,
$$
where $\q^+$ is a four-component anticommuting spinor
variable, and $m=0,1,2,3$. 

The concept of harmonic superspace was 
originally developed for 4D, $\cN=2$ 
supersymmetric theories including 
supergravity \cite{GIKOS},
and by now it has become a textbook subject\footnote{The 
book \cite{GIOS} contains a list of relevant publications
in the context of  harmonic superspace.}
\cite{GIOS}. Actually it can be argued that  
harmonic superspace is a natural framework for 
all supersymmetric theories with eight supercharges, 
both at the classical and quantum levels.
In the case of four space-time dimensions, 
probably the main objection to this approach 
was the issue  that theories in harmonic superspace 
are often difficult to reduce to $\cN=1$ superfields
(the kind of reduction which brane-world practitioners 
often need).
But this objection has been lifted since the 
advent and subsequent perfection of 4D, $\cN=2$ projective 
superspace \cite{projective0,projective} 
which  allows a nice reduction to $\cN=1$ superfields
and which appears to be  a truncated 
version of the harmonic superspace \cite{Kuzenko:1998xm}. 

What is the difference between harmonic superspace and projective 
superspace? In five space-time dimensions (to be concrete),
they make use of the same supermanifold
${\mathbb R}^{5|8} \times S^2$, with 
${\mathbb R}^{5|8} $ the conventional 5D simple superspace. 
In harmonic superspace, one deals with so-called 
{\it Grassmann analytic} (also known as {\it twised chiral}) superfields 
that are chosen to be smooth tensor  fields on $S^2$. 
In projective superspace, one also deals with 
Grassmann analytic superfields that are holomorphic functions
on an open subset of $S^2$. It is clear that the harmonic superspace 
setting is more general. Actually, many results originally 
obtained in projective superspace can be reproduced 
from harmonic superspace by applying special truncation procedures
 \cite{Kuzenko:1998xm}.  The remarkable features of 
projective superspace are that (i) the projective supermultiplets
are easily represented as a direct sum of standard 4D, $\cN=1$ 
superfields; (ii) this approach provides simple rules
to construct low-energy effective actions that are easily expressed in 
terms  of 4D, $\cN=1$  superfields. Of course, one could wonder why 
both harmonic and projective superspaces should be introduced?
The answer is that, in many respects, they are complementary 
to each other. (This is analogous to the relation between 
the theorems of existence of solutions for differential equations
and concrete techniques to solve such equations.)

To avoid technicalities, in this paper we do not consider 
5D superfield supergravity at all, and concentrate only
on developing a 5D simple superspace approach to 
globally supersymmetric gauge theories. One of our 
main objectives is to demonstrate that 5D superspace 
may be useful, even in the context of 4D effective theories 
with an extra dimension.
Here we develop  manifestly 5D supersymmetric techniques
which, on the one hand, allow us to construct 
many  of the 5D supersymmetric models originally developed 
within the ``hybrid''  formulation. One the other hand, 
these techniques make it possible to construct very interesting supersymmetric
nonlinear sigma-models whose construction is practically beyond the
scope of the ``hybrid''  formulation.  Examples of such 
5D supersymmetric sigma-models
are constructed for the  first time below.
We therefore believe that the paper should be of some interest to
both superspace experts and newcomers. 

It is worth saying a few words about the global structure of this paper.
We are aiming at (i) elaborating 5D off-shell matter supermultiplets
and their superfield descriptions; 
(ii) developing various universal procedures to construct  manifestly 
5D supersymmetric action functionals, and then applying them 
to specific supermultiplets;
(iii)  elaborating on techniques to reduce such super-actions 
to 4D, $\cN=1$ superfields.
New elements of 5D superfield formalism are introduced
only if they are  essential for further consideration. 
For example, the Yang-Mills off-shell supermultiplet can be realized 
in 5D conventional superspace in terms of constrained 
superfields. In order to solve the constraints, however, one has to 
introduce the concept of harmonic superspace.

This paper is organized as follows:
In section \ref{section:two} we describe, 
building on earlier work \cite{HL,Z},
the 5D Yang-Mills supermultiplet 
and its salient properties, both 
in the  conventional and harmonic superspaces.
We also describe several off-shell realizations 
for the 5D hypermultiplet. 
In section \ref{section:Actions} we present two procedures to construct
5D manifestly supersymmetric actions for multiplets
with and without intrinsic central charge, and give several 
examples. 
Section \ref{section:four} is devoted to 5D supersymmetric Chern-Simons 
theories. Their harmonic superspace actions are given 
in a new form, as compared with \cite{Z},
which allows a simple reduction to  the projective superspace. 
We also uncover the 5D origin
for the superfield constraints describing the so-called
4D, $\cN=2$ nonlinear vector-tensor multplet.
In section \ref{section:five}, some of the results developed
in the previous sections
are  reduced to a ``hybrid'' formulation 
which keeps manifest only 4D, $\cN=1$ super 
Poincar\'e symmetry. 
Section \ref{section:six} introduces 5D simple projective 
superspace and projective multiplets.
Here we also present two families of 5D  
off-shell supersymmetric nonlinear sigma-models 
which are formulated, respectively,  
in terms of a (i) 5D tensor multiplet; (ii) 5D polar mutiplet.
Section \ref{section:seven} deals with the vector multiplet in projective superspace.
A brief conclusion is given in section \ref{section:eight}.
This paper also includes three technical appendices.
Appendix \ref{section:A} contains our 5D notation and conventions, 
inspired by those in \cite{WB,BB}, as well as some important
identities. 
Appendix \ref{section:B} is devoted to a review
of the well-known one-to-one
correspondence between smooth tensor fields on 
$S^2 = {\rm SU}(2)/{\rm U}(1)$ and
smooth scalar functions over SU(2) with definite U(1) charges.
Finally, in appendix \ref{section:C} we briefly demonstrate, 
mainly following  \cite{Kuzenko:1998xm}, 
how to derive the projective superspace action
(\ref{integral-projective}) from 
the harmonic superspace action (\ref{integral-analyt}).

\sect{5D Supersymmetric Matter}
\label{section:two}

\subsection{Vector multiplet in conventional superspace}
To describe a Yang-Mills supermultiplet  in 5D simple
superspace ${\mathbb R}^{5|8}$ parametrized  
by  coordinates  $ z^{\hat A} = (x^{\hat a},  \q^{\hat \a}_i )$
we introduce  gauge-covariant derivatives\footnote{Our
notation and conventions are collected in Appendix A.}
\be
\cD_{\hat A} = ( \cD_{\hat a}, \cD_{\hat \a}^i ) 
= D_{\hat A} + {\rm i}\, \cV_{\hat A} (z) ~,
\qquad 
[ \cD_{\hat A} \, ,\, \cD_{\hat B} \} = T_{\hat A \hat B}{}^{\hat C} \, 
\cD_{\hat C}  + C_{\hat A \hat B} \, \D
+{\rm i} \, \cF_{\hat A \hat B} 
~,
\ee 
with $D_{\hat A} = ( \pa_{\hat a}, D_{\hat \a}^i ) $
the flat covariant derivatives obeying 
the anti-commutation relations (\ref{flat}),  
$\D$ the central charge,  
and $  \cV_{\hat A} $ the gauge connection taking its 
values in the Lie algebra of the gauge group.
The connection is chosen to be inert under the central 
charge transformations, 
$[\D \,,\cV_{\hat A} ] =0$.
The operators $\cD_{\hat A}$ possess 
the following gauge transformation law
\be
\cD_{\hat A} ~\mapsto ~ {\rm e}^{{\rm i} \t(z)} \, \cD_{\hat A}\,
{\rm e}^{-{\rm i} \t(z)}~, \qquad 
\t^\dagger = \t ~, \qquad [ \D\, , \t ]=0~,
\label{tau}
\ee
with the gauge parameter $\t(z)$ being arbitrary modulo 
the reality condition imposed. 
The gauge-covariant derivatives are required   
to obey some constraints \cite{HL} such that 
\bea
\{\cD^i_{\hat \a} \, ,  \cD^j_{\hat \b} \} &= &-2{\rm i} \,
\ve^{ij}\,
\Big( (\G^{\hat c} ){}_{\hat \a \hat \b} \, \cD_{\hat c} 
+ \ve_{\hat \a \hat \b} \,( \D +{\rm i}\, \cW ) \Big)~, 
\qquad \big[ \cD^i_{\hat \a} \, ,  \D \big] =0~,
\non \\
\big[ \cD_{\hat a} \, ,  \cD^j_{\hat \b} \big] &=& 
{\rm i}\,  (\G_{\hat a} ){}_{\hat \b}{}^{\hat \g} \,
\cD^j_{\hat \g} \cW~, \qquad 
\big[ \cD_{\hat a} \, , \cD_{\hat b} \big]
= -{1 \over 4}  (\S_{\hat a \hat b})^{\hat \a \hat \b} \,
\cD^i_{\hat \a} \cD_{\hat \b i}  \cW
= {\rm i}\, \cF_{\hat a\hat b}~, 
\label{SYM-algebra}
\eea
 with the matrices $\G_{\hat a} $ and $\S_{\hat a \hat b}$ 
defined in Appendix A.
Here the field strength $\cW$ is hermitian, 
$\cW^\dag = \cW$, and obeys the Bianchi identity
(see e.g. \cite{Z})
\be
\cD^{(i}_{\hat \a} \cD_{\hat \b }^{j)}  \cW
= {1 \over 4} \ve_{\hat \a \hat \b} \,
\cD^{\hat \g (i} \cD_{\hat \g }^{j)}  \cW~, 
\label{Bianchi1}
\ee
and therefore
\be
\cD^{(i}_{\hat \a} \cD_{\hat \b }^{j}  \cD_{\hat \g }^{k)} \cW
= 0~.
\label{Bianchi1.5}
\ee

The independent component fields contained in $\cW$ are:
\bea
\vf = \cW {\double |} ~, \quad
{\rm i} \,\J^i_{\hat \a} =  \cD_{\hat \a }^{i}  \cW {\double |}~, 
\quad -4 {\rm i} \,F_{\hat \a \hat \b} = 
\cD^i_{( \hat \a} \cD_{\hat \b) i}  \cW {\double |}~, \quad
-4{\rm i} \, X^{i j} = \cD^{\hat \a (i} \cD_{\hat \a }^{j)}  \cW {\double |}~.
\label{W-components}
\eea
Here and in what follows, $U{\double |}$ denotes the $\q$-independent
component of a superfield $U(x,\q)$.
It is worth noting that
\be
F_{\hat a \hat b} = \cF_{\hat a \hat b} {\double |}~.
\ee

\subsection{Vector multiplet in harmonic superspace}
The most elegant way to solve the constraints 
encoded in the algebra (\ref{SYM-algebra}) is to use 
the concept of harmonic superspace 
originally developed for 4D, $\cN=2$ 
supersymmetric theories \cite{GIKOS,GIOS}
(related ideas appeared in   \cite{RS}).
In this approach, the conventional superspace ${\mathbb R}^{5|8}$ 
is embedded into  
 ${\mathbb R}^{5|8} \times  S^2$, 
where the two-sphere $ S^2 = {\rm SU}(2) /{\rm U}(1)$
is parametrized by so-called
harmonic $u_i{}^- $ and  $u_i{}^+$,  that is group elements
\be
(u_i{}^- \, , u_i{}^+) \in {\rm SU}(2) ~, 
\quad u_i^+ = \ve_{ij}\, u^{+j} ~, \quad
(u^{+i})^* = u^-_i ~, \quad 
u^{+i} u^-_i =1~.
\ee
As is well-known, tensor fields over $ S^2$ are 
in  a one-to-one correspondence with functions
over SU(2) possessing definite harmonic U(1) charge
(see \cite{Kuzenko:1998xm} for a review).
A function $\J^{(p) }(u)$ is said to have 
harmonic U(1) charge $p$ if 
\bea
\J^{(p) }({\rm e}^{ {\rm i}\a} u^+ , {\rm e}^{ -{\rm i}\a} u^-)
= {\rm e}^{ {\rm i}p \a} \,
\J^{(p) }(u^+ ,  u^-)~,  \qquad |{\rm e}^{ {\rm i}\a} |=1~.
\eea
Such functions, extended to the whole harmonic superspace
${\mathbb R}^{5|8} \times  S^2$, that is $\J^{(p) }(z,u)$, 
are called harmonic superfields. 
Introducing the harmonic derivatives \cite{GIKOS}
\bea
D^{++}=u^{+i}\frac{\partial}{\partial u^{- i}} ~,\quad
D^{--}=u^{- i}\frac{\partial}{\partial u^{+ i}} ~,\quad
D^0&=&u^{+i}\frac{\partial}{\partial u^{+i}}-u^{-i}
\frac{\partial}{\partial u^{-i}} ~,\non \\
{[}D^0,D^{\pm\pm}]=\pm 2D^{\pm\pm}~, \qquad  [D^{++},D^{--}]&=&D^0~,
\label{5}
\eea
one can see that $D^0$ is the operator of harmonic U(1) charge,
$D^0 \, \J^{(p) }(z,u)=p\,\J^{(p) }(z,u)$.
Defining
\be
{\cal D}_{\bA}\equiv({\cal D}_{\hat A},{\cal D}^{++},{\cal D}^{--},
{\cal D}^0)~, \qquad {\cal D}^{\pm\pm}=D^{\pm\pm}~, \qquad
{\cal D}^0=D^0~,
\label{cd-tau}
\ee
one observes that the operators ${\cal D}_{\bA}$ possess the
same transformation law (\ref{tau}) 
as ${\cal D}_{\hat A}$.

Introduce a new basis for the spinor covariant derivatives:
 $\cD^+_{\hat \a} = \cD^i_{\hat \a} \,u^+_i $ 
and $\cD^-_{\hat \a} = \cD^i_{\hat \a} \,u^-_i $.
Then, eq.  (\ref{SYM-algebra}) leads to
\bea
\{\cD^+_{\hat \a} \, ,  \cD^+_{\hat \b} \} &=&0~,
\qquad \quad ~~ 
\big[\cD^{++}\, , \cD^+_{\hat \a} \big] =0~,
 \non \\
\{\cD^+_{\hat \a} \, ,  \cD^-_{\hat \b} \} &=&2{\rm i}\,
\Big( \cD_{\hat \a \hat \b}  
+ \ve_{\hat \a \hat \b} \,( \D +{\rm i}\, \cW ) \Big)
~,  \\
\big[\cD^{++}\, , \cD^-_{\hat \a} \big] &=&\cD^+_{\hat \a}~, 
\qquad \quad
\big[\cD^{--}\, , \cD^+_{\hat \a} \big] =\cD^-_{\hat \a}~. 
\non
\eea
In  harmonic superspace,
the integrability condition $\{\cD^+_{\hat \a} \, ,  \cD^+_{\hat \b} \} =0$
 is solved by 
\be
\cD^+_{\hat \a} = {\rm e} ^{-{\rm i} \Omega } \, D^+_{\hat \a} \,
{\rm e} ^{{\rm i} \Omega }~, 
\ee
for some Lie-algebra valued harmonic superfield $\Omega = \Omega (z, u)$ 
of vanishing harmonic U(1) charge, $D^0\, \Omega= 0$. This superfield
is called the bridge. The bridge possesses a richer 
gauge freedom than the original $\t$-group (\ref{tau})
\be
{\rm e} ^{{\rm i} \Omega (z,u)}
~\mapsto ~ 
{\rm e}^{{\rm i} \lambda(z,u)} \,  {\rm e} ^{{\rm i} \Omega (z,u)}  \,
{\rm e}^{-{\rm i} \t(z)}~, \qquad 
D^+_{\hat \a} \,\lambda =0~, 
\qquad [ \D\, , \lambda ]=0~.
\label{O-trans} 
\ee
The $\lambda$- and $\t$-transformations
generate, respectively, the so-called
$\lambda$- and $\t$-groups. 

One  can now define {\it covariantly
analytic} superfields constrained by
\be
{\cal D}^+_{\hat \alpha}\Phi^{(p)}=0~.
\label{10}
\ee
Here $\Phi^{(p)}(z,u)$ carries U(1)-charge $p$, 
$D^0\Phi^{(p)}=p\Phi^{(p)}$, and can be represented as follows
\be
\Phi^{(p)}={\rm e}^{-{\rm i}\Omega}\phi^{(p)} ~,\qquad
D^+_{\hat \alpha}\phi^{(p)}=0~,
\label{11}
\ee
with $\phi^{(p)}(\zeta)$ 
being an {\it analytic} superfield -- that is, a field  
over  the so-called  {\it analytic subspace} 
of the harmonic superspace  parametrized by 
\be
\zeta \equiv\{\bm{x}^{\hat a},\theta^{+\hat \alpha}, u^+_i, u^-_j \}~,
\label{analytsub}
\ee
where 
\be
\bm{x}^{\hat a} = x^{\hat a} + {\rm i} \,(\G^{\hat a})_{\hat \b \hat \g} \, 
\q^{+\hat \b} \q^{-\hat \g}~,\qquad
\theta^\pm_{\hat \a} = \theta^i_{\hat \a}\,u^\pm_i~.
\ee
In particular, the gauge parameter $\lambda$ in (\ref{O-trans}) is
an unconstrained analytic  superfield of vanishing harmonic U(1) 
charge, $D^0 \lambda =0$.  It is clear that
the superfields $\Phi^{(p)}$ and $\phi^{(p)}$ describe 
the same matter multiplet but in different frames
(or, equivalently,  representations), and they transform 
under the $\tau$-  and $\lambda$-gauge groups, respectively.
\be 
\Phi^{(p)}(z,u) ~\mapsto ~
{\rm e}^{{\rm i} \t(z)}\, \Phi^{(p)}(z,u)~, \qquad 
\phi^{(p)}(z,u) ~\mapsto ~
{\rm e}^{{\rm i} \lambda(z,u)}\, \phi^{(p)}(z,u)~.
\ee

By construction, the analytic subspace (\ref{analytsub}) is closed under
the  supersymmetry transformations. 
Unlike the chiral subspace,
it is real with respect to the generalized conjugation 
(often called the smile-conjugation)
$\, \breve{} \, 
$ \cite{GIKOS} defined to be the composition of the complex conjugation
(Hermitian conjugation in the case of Lie-algebra-valued
superfields)
with the operation ${}^\star$ acting on the harmonics only
\bea
(u^+_i)^\star = u^-_i~, \qquad
(u^-_i)^\star = - u^+_i \quad \Rightarrow \quad
(u^{\pm}_i)^{\star \star} = - u^{\pm}_i~,
\eea
hence
\be
(u^{+i}) \,\breve{{}} = - u^+_i  \qquad \quad (u^-_i)\,\breve{{}} = u^{-i}\;.
\label{breve}
\ee
The analytic superfields of even U(1) charge
may therefore be chosen to be real.
In particular, the bridge $\Omega$ and the gauge parameter $\lambda$ 
are real.
 
The covariant derivatives in the $\lambda$-frame are obtained 
from those in the $\tau$-frame, eq. (\ref{cd-tau}), by applying the transformation 
\be
{\cal D}_{\bA} \quad \mapsto \quad 
{\rm e}^{{\rm i}\Omega}\,  \cD_{\bA}\,{\rm  e}^{-{\rm i}\Omega}~.
\ee 
Then, the gauge transformation 
of the covariant derivatives becomes
\be
\cD_{\bA} ~\mapsto ~ {\rm e}^{{\rm i} \lambda(\z)} \, \cD_{\bA}\,
{\rm e}^{-{\rm i} \lambda(\z)}~, \qquad 
\breve{\lambda} = \lambda ~, \qquad [ \D\, , \lambda ]=0~.
\label{lambda}
\ee
In the $\lambda$-frame, the spinor covariant 
derivatives  $\cD^+_{\hat \a} $ coincide with the flat ones, 
$ \cD^+_{\hat \a} = D^+_{\hat \a} $, while the harmonic covariant derivatives
acquire connections, 
\be
\label{eqn:HarmConnex}
\cD^{\pm \pm} = D^{\pm \pm} +{\rm i} \, \cV^{\pm \pm}~.
\ee 
 The {\it real} connection $\cV^{++}$ is seen to be an 
analytic superfield, $D^+_{\hat \a} \cV^{++}=0$,
of harmonic U(1) charge plus two,
$D^0 \,\cV^{++} =2 \cV^{++}$.
The other harmonic connection $\cV^{--}$
turns out to be uniquely determined 
in terms  of $\cV^{++}$ using 
the zero-curvature condition
\be
[\cD^{++}\,, \cD^{--} ] = D^0 \quad 
\Longleftrightarrow \quad
D^{++} \cV^{--} - D^{--} \cV^{++}  
+{\rm i} \,  [\cV^{++}\,, \cV^{--} ] =0~,
\label{zero-curv}
\ee
as demonstrated in \cite{Z2}. 
The result is
\bea
\cV^{--} (z,u)=
\sum \limits_{n=1}^{\infty}  (- {\rm i} )^{n+1} \,
\int  {\rm d} u_1 \dots {\rm d} u_n \,
\frac{
\cV^{++} (z,u_1) \,\cV^{++} (z,u_2) 
\cdots \cV^{++} (z,u_n)
 }
{ (u^+ u^+_1) (u^+_1 u^+_2) \dots (u^+_n u^+ ) }~,
\label{V--}
\eea
with  $ (u^+_1 u^+_2) = u_1^{+i} u^+_{2}{}_i$, 
and the harmonic distributions on the right of
(\ref{V--}) defined, e.g.,  in \cite{GIOS}.
Integration over the group manifold 
SU(2) is normalized according to \cite{GIKOS}
\be 
 \int {\rm d}u \, 1 = 1~,\qquad 
 \int {\rm d}u \, u^+_{(i_1} \cdots u^+_{i_n}\,
u^-_{j_1} \cdots u^-_{j_m)} =0~, 
\quad n+m >0~.
\ee

As far as the connections $\cV^-_{\hat \a} $ 
and $\cV_{\hat a}$ are concerned, they can be  expressed in terms 
of $\cV^{--}$ with the aid of the (anti-)commutation relations
\bea 
\big[\cD^{--}\, , \cD^+_{\hat \a} \big] &=&\cD^-_{\hat \a}~, 
\qquad 
\{\cD^+_{\hat \a} \, ,  \cD^-_{\hat \b} \} = 2{\rm i} \,
\Big(  
\cD_{\hat \a \hat \b} 
+ \ve_{\hat \a \hat \b} \,( \D +{\rm i}\, \cW_{\lambda} ) \Big)~. 
\eea
In particular, one obtains 
\be
\cW_{\lambda} = \frac{\rm i}{8} ({\hat D}^+)^2 \, \cV^{--}~, 
\qquad 
({\hat D}^+)^2 = D^{+\,\hat \a} D^+_{\hat \a}~,
\label{W1}
\ee
where $\cW_{\lambda} $ stands for the field strength in 
the $\lambda$-frame.
Therefore, $\cV^{++}$ is the single unconstrained 
analytic prepotential of the theory.
With the aid of  (\ref{zero-curv}) one can obtain  
the following useful expression 
\bea
\cW = \frac{\rm i}{8} \int {\rm d}u \, 
 ({\hat D}^-)^2 \, \cV^{++}
 ~+~ O \Big((\cV^{++})^2\Big)~. 
 \label{W2}
  \eea
 In the Abelian case, only the first term on the right survives.
In what follows, we do not distinguish between
$\cW$ and $\cW_{\lambda}$.

With the notation 
$({\hat \cD}^+)^2 = \cD^{+\,\hat \a} \cD^+_{\hat \a}$,
the Bianchi identity (\ref{Bianchi1})
takes the form 
\be
\cD^+_{\hat \a} \cD^+_{\hat \b }  \cW
= {1 \over 4} \ve_{\hat \a \hat \b} \,
({\hat \cD}^+)^2  \cW \quad 
\Rightarrow \quad 
\cD^+_{\hat \a} \cD_{\hat \b }^+  \cD_{\hat \g }^+ \cW
= 0~.
\label{Bianchi2}
\ee

Using the Bianchi identity (\ref{Bianchi2}), 
one can readily construct a covariantly 
analytic descendant of $\cW$
\be
-{\rm i} \, \cG^{++} = 
\cD^{+ \hat \a} \cW \, \cD^+_{\hat \a} \cW 
+{1 \over 4} \{\cW \,, 
({\hat \cD}^+)^2 \cW \}~, \qquad 
\cD^+_{\hat \a} \cG^{++} = 
 \cD^{++} \cG^{++} =0~.
\label{YML}
\ee

\subsection{Vector multiplet in components}
The gauge freedom 
\be
-\d \cV^{++} = \cD^{++} \lambda  
=  D^{++} \lambda  +{\rm i} \,[ \cV^{++}\,, \lambda] 
\label{lambda2}
\ee
can be used to choose a Wess-Zumino gauge
of the form 
\bea 
\cV^{++}(\bm{x}, \q^+,u) &=& {\rm i} \,(\hat{\q}^+)^2 \,
\vf(\bm{x})  - {\rm i}\,
\q^+ \G^{\hat m} \q^+ \,A_{\hat m}(\bm{x})
+4(\hat{\q}^+)^2 \q^{+ \hat \a} \, u^-_i \,
\J^i_{\hat \a}(\bm{x}) \non \\
&-&{3\over 2} \,(\hat{\q}^+)^2 (\hat{\q}^+)^2\,u^-_i u^-_j \, 
X^{ij}(\bm{x})~, 
\label{WZ-V++}
\eea
where 
\be
(\hat{\q}^+)^2 = \q^{+\hat \a} \q^+_{\hat \a}~, 
\qquad 
 \q^+ \G^{\hat m} \q^+ =  \q^{+ \hat \a} \,
( \G^{\hat m})_{\hat \a}{}^{\hat \b}  \,
\q^+{}_{ \hat \b} 
=-( \G^{\hat m})_{\hat \a \hat \b}  \,\q^{+ \hat \a} 
\q^{+ \hat \b} ~.
\ee
In this gauge, the expression (\ref{V--}) 
simplifies considerably
\bea
\cV^{--}(z,u) &=& 
\int  {\rm d} u_1 \,
\frac{
\cV^{++} (z,u_1)
 }
{ (u^+ u^+_1)^2 }
+{ {\rm i} \over 2} \int  {\rm d} u_1\,  {\rm d} u_2 \,
\frac{
[\cV^{++} (z,u_1)\,, \cV^{++} (z,u_2)] 
 }
{ (u^+ u^+_1) (u^+_1 u^+_2)  (u^+_2 u^+ ) }
\non \\
&+& \mbox{terms of third- and fourth-order in}~ \cV^{++}~.
\label{WZ-V--}
\eea
Here the explicit form of the cubic and quartic terms 
is not relevant for our consideration.
One of the important properties of the Wess-Zumino 
gauge is 
\be
\cD_{\hat m} {\double |} \equiv  \pa_{\hat m} + 
{\rm i}\,  \cV_{\hat m} {\double |} = \pa_{\hat m} + 
{\rm i} \, A_{\hat m} (x) ~.
\label{two-cov-der}
\ee
The component fields of $\cW$ and $\cV^{++}$
can be related to each other 
using the identity
\be
F^+_2 = (u^+_1 u^+_2) \, F^-_1   - (u^-_1 u^+_2) \, F^+_1~, 
\qquad \quad
F^\pm = F^i \,u^\pm_i~,
\ee
and the analyticity of $\cV^{++}$.
(The latter property implies, for instance, 
 $D^+ \cV^{++} (z,u_1)
= -(u^+ u^+_1)\, D^-_1 \cV^{++} (z,u_1)$.)
Thus one gets 
\bea
\label{WZ-gauge}
\cW {\double |} = \frac{\rm i}{8} ({\hat D}^+)^2 \, \cV^{--}{\double |}
=  \frac{\rm i}{8} \int  {\rm d} u_1 \,({\hat D}^-_1)^2\,
\cV^{++} (z,u_1){\double |} &=&\vf (x) ~, \non \\
\cD^+_{\hat \a} \cW {\double |} = - { {\rm i} \over 8}
 \int  {\rm d} u_1 \,(u^+ u^+_1)\,
D^-_{1 \,\hat \a}({\hat D}^-_1)^2\, \cV^{++} (z,u_1){\double |} 
&=&{\rm i}\, \J^i_{\hat \a} (x) \, u^+_i~, \\
({\hat \cD}^+)^2 \cW {\double |} = { {\rm i} \over 8}
 \int  {\rm d} u_1 \,(u^+ u^+_1)^2\,
({\hat D}^-_1)^2\,({\hat D}^-_1)^2\,
\cV^{++} (z,u_1){\double |} 
&=&-4{\rm i}\, X^{ij} (x) \, u^+_i u^+_j~.
 \non
\eea
${}$Finally, eq. (\ref{two-cov-der}) implies that
the component field 
$F_{\hat \a \hat \b} = F_{( \hat \a \hat \b )} $ 
in (\ref{W-components})  
is (the bispinor form of) the gauge-covariant 
field strength $F_{\hat m \hat n}$
generated by the  gauge field
$A_{\hat m}$.

\subsection{Fayet-Sohnius hypermultiplet}
${}$Following the four-dimensional 
$\cN=2$ supersymmetric construction
due to Fayet and Sohnius \cite{Fayet,Sohnius}, 
an off-shell hypermultiplet with intrinsic 
central charge, which is coupled to the
Yang-Mills  supermultiplet,
can be described by a superfield ${\bm q}^i (z)$ 
and its conjugate ${\bar {\bm q}}_i (z)$,
${\bar {\bm q}}_i = ({\bm q}^i )^\dagger$,
subject to the constraint 
\be
\cD^{(i}_{\hat \a} \, {\bm q}^{j)} =0~. 
\label{FSh}
\ee
Introducing ${\bm q}^+(z,u) = {\bm q}^i (z) \,u^+_i$
and $\breve{{\bm q}}{}^+(z,u) = 
-{\bar {\bm q}}^i (z) \,u^+_i$, 
the constraint (\ref{FSh}) can be rewritten in the form 
\be 
\cD^+_{\hat \a} \,{\bm q}^+ = \cD^+_{\hat \a} \,
\breve{{\bm q}}{}^+ =0~, 
\qquad 
\cD^{++} \, {\bm q}^+ =\cD^{++} \, \breve{\bm q}{}^+ =0~.
\ee
Thus ${\bm q}^+$ is a constrained  analytic superfield.
Using the algebra of gauge-covariant derivatives, 
the constraints can be shown to imply\footnote{By 
analogy with the four-dimensional case \cite{BBKO},
the operator $ {\stackrel{\frown}{\Box}}$ can be  called 
the covariant analytic d'Alembertian. 
Given a covariantly analytic superfield $\Phi^{(q)}$, 
the identity 
$ {\stackrel{\frown}{\Box}} \Phi^{(q)} 
=-{1 \over 64} ({\hat \cD}^+)^2 ({\hat \cD}^+)^2 
(\cD^{--})^2  \Phi^{(q)} $ holds, and therefore
${\stackrel{\frown}{\Box}} $
preserves analyticity.}   
\bea
&&{\stackrel{\frown}{\Box}} \,{\bm q}^+ =0~, \\
&&{\stackrel{\frown}{\Box}} = 
\cD^{\hat a} \cD_{\hat a} 
+( {\cal D}^{+ \hat \a}\cW)\,{\cD}^-_{\hat \a}
-{ 1 \over 4} 
({\hat \cD}^{+ \hat \a} \cD^+_{\hat \a} \cW )\, \cD^{--}
+\frac{1}{8}[{\cD}^{+ \hat \a},{\cal D}^-_{\hat \a}] \,\cW
+(\D +{\rm i}\, \cW)^2 ~. \non
 \eea
Therefore,  the requirement of a constant central charge,
$\D \, {\bm q}^+ = m \,{\bm q}^+$,   
with $m$ a constant mass parameter,
is equivalent to an equation of motion 
for  the hypermultiplet.

Independent component fields of ${\bm q}^i (z)$ 
can be chosen as 
\bea
C^i= {\bm q}^i {\double |}~, \qquad
\lambda_{\hat \a}={{\rm i}\over \sqrt{8}}\,\cD^i_{\hat \a} \,{\bm q}_i
{\double |}~,
\qquad 
F^i=\D {\bm q}^i {\double |}~.
\label{FS-components}
\eea
All other components can be related to these and their derivatives. 
${}$For example, 
\bea
(\hat \cD^-)^2 {\bm q}^+&=&8{\rm i}\, 
\D {\bm q}^- - 8\cW\, {\bm q}^-~,\cr
\cD^-_{\hat \a} \D {\bm q}^+&=&
\cD_{\hat \a}{}^{\hat \b}
 \cD^+_{\hat \b} {\bm q}^-+ 
{\rm i} \, \cW\,\cD^+_{\hat \a} {\bm q}^-
+2{\rm i}\, (\cD^+_{\hat \a} \cW)\, {\bm q}^-~,\cr
(\hat \cD^-)^2 \D {\bm q}^+&=&
-8{\rm i}\, \cD^{\hat a} \cD_{\hat a}  {\bm q}^-
+8\cW\, \D {\bm q}^-
+8{\rm i}\, \cW^2 \,{\bm q}^-
+4{\rm i}\, ( \cD^{-\hat \a} \cW)\, \cD^+_{\hat \a} {\bm q}^-\cr
&&+2{\rm i}\, ( \cD^{-\hat \a} \cD^+_{\hat \a} \cW)\, {\bm q}^-
-2{\rm i}\, ( \cD^{-\hat \a} \cD^-_{\hat \a}
\cW)\, {\bm q}^+~.
\eea

\subsection{Off-shell hypermultiplets without central charge}
One of the main virtues of the harmonic superspace approach 
\cite{GIKOS} is that it makes possible 
an off-shell formulation 
for a charged hypermultiplet 
(transforming in an arbitrary representation of the gauge group)  
without central charge.
Such a  $q^+$-{\it hypermultiplet}
is described by an unconstrained analytic superfield 
$q^+(z,u)$ and its conjugate $\breve{q}^+(z,u)$, 
\be
\cD^+_{\hat \a} \,q^+ =0~, 
\qquad 
\D \, q^+ =0~.
\label{q-hyper}
\ee
In this approach, the requirement 
that $q^+$ be  a holomorphic spinor field over the two-sphere, 
$\cD^{++}\, q^+=0$, is equivalent 
to an equation of motion.\footnote{The equation of motion 
for the massless Fayet-Sohnius hypermultiplet, 
which is characterised by the kinematic constraint 
$\cD^{++}\,{\bm q}^+=0$, can be shown to be 
$\D \, {\bm q}^+ =0$, if the dynamics is generated
by the Lagrangian (\ref{FS-Lagrangian}) with $m=0$.}
The harmonic dependence of the $q^+$-hypermultiplet
is non-trivial. 
One can represent $q^+(z,u)$ by a convergent Fourier series
of the form (\ref{smoothfunction}) with $p=1$. 
The corresponding Fourier 
coefficients $q^{i_1 \cdots i_{2n+1}} (z)$,
where $n=0,1, \dots$, obey some constraints 
that follow from the analyticity condition in (\ref{q-hyper}).

Given a hypermultiplet that transforms in a real representation of the gauge 
group, it can be described by a real anaytic superfied $\omega(z,u)$, 
\be
\cD^+_{\hat \a} \,\omega =0~, 
\qquad 
\D \, \omega =0~,
\label{o-hyper}
\ee
called the $\omega$-{\it hypermultiplet} \cite{GIKOS}.
The gauge parameter $\lambda$ in (\ref{lambda}) is of this superfield type. 
It is then clear that the
$\omega$-hypermultiplet  can be used, for instance, 
to formulate a gauge-invariant St\"uckelberg description 
for massive vector multiplets.

\sect{Supersymmetric Actions}
\label{section:Actions}

In the case of vanishing central charge, $\D=0$, 
it is easy to construct manifestly supersymmetric actions
within the harmonic superspace approach \cite{GIKOS}.
Given a scalar harmonic superfield $L(z,u)$
and a scalar analytic superfield $L^{(+4)} (\z)$, 
supersymmetric actions are:
\bea  
S_{\rm H} &=& 
\int {\rm d}^5x  \,
{\rm d}^8\q \, 
 {\rm d}u \, L 
= \int {\rm d}^5 x \,
{\rm d} u \, 
(\hat{D}^-)^4  \,(\hat{D}^+)^4\,
L \double{\Big|}~,~ 
\label{integral-full}\\
S_{\rm A}&=&   \int {\rm d} \z^{(-4)} \, L^{(+4)}
=\int {\rm d}^5 x \,
{\rm d} u \, 
(\hat{D}^-)^4  \,L^{(+4)}\double{\Big|}
~, \qquad \cD^+_{\hat \a}  L^{(+4)} =0~,
\label{integral-analyt}
\eea
where 
\be
(\hat{D}^\pm)^4 = -{1\over 32}  (\hat{D}^\pm)^2 
\,  (\hat{D}^\pm)^2~.
\label{D4+-}
\ee 
As follows from (\ref{integral-full}), any integral 
over the full superspace can be reduced to 
an integral over the analytic subspace, 
\be
\int {\rm d}^5x\, 
{\rm d}^8\q \, 
 {\rm d}u \, L 
=\int {\rm d} \z^{(-4)} \, L^{(+4)}~, 
\qquad 
 L^{(+4)} = (\hat{D}^+)^4\,L~.
 \ee

The massless $q^+$-hypermultiplet action \cite{GIKOS} is 
\be
\label{eqn:OffShellq+}
S= - \int  {\rm d} \zeta^{(-4)}\,
\breve{q}{}^+ \cD^{++}q^+ ~.
\ee
This action also describes a massive hypermutliplet 
if one assumes that (i) the gauge group is $G \times {\rm U(1)}$, 
and (ii) the  U(1) gauge field $\cV^{++}_0$ possesses 
a constant field strength $\cW_0= {\rm const}$,  
$|\cW_0|=m$, see \cite{vev} 
for more details.\footnote{A different approach to formulate 
massive hypermultiplets was proposed in \cite{Ohta1}.}

Similarly to the chiral scalar in 4D, $\cN=1$ supersymmerty, 
couplings for the $q^+$-hypermultiplet  are easy to construct. 
${}$For example, one can consider a Lagrangian of the form
\bea
L^{(+4)} = -\breve{q}{}^+ \cD^{++}q^+ + \lambda\, (\breve{q}{}^+ q^+)^2
+ \breve{q}{}^+ \Big\{ \s_1 \, (\hat{\cD}^+)^2 \cW 
+ {\rm i} \s_2 \,\cG^{++} \Big\} q^+ ~,
\eea
with the quartic self-coupling first introduced in \cite{GIKOS}.
Consistent couplings for the Fayet-Sohnius hypermultiplet 
are much more restrictive, as a result of a non-vanishing 
intrinsic central charge. 

\subsection{Four-derivative vector multiplet action} 
As another example of supersymemtric action,
we consider four-derivative couplings that may occur
in low-energy effective actions for a U(1) vector multiplet.
\bea
S_{\rm four-deriv} &=& 
\int  {\rm d} \zeta^{(-4)}\,
\cG^{++} \Big\{ \k_1 \, \cG^{++} + {\rm i}\, \k_2\,
({\hat D}^+)^2 \cW \Big\} 
+ \int {\rm d}^5x  \,
{\rm d}^8\q \, 
H(\cW) ~,
\eea
with $\k_{1,2} $ coupling constants, 
the analytic superfield $\cG^{++} $ 
given by (\ref{YML}), and $H(\cW)$ an arbitrary 
function. The third term on the right is a natural generalization 
of the four-derivative terms in 4D, $\cN=2$ supersymmetry
first introduced in \cite{Hen}.

\subsection{Multiplets with intrinsic central charge}
${}$For 5D off-shell supermultiplets with $\D \neq 0$, 
the construction of supersymmetric actions is based 
on somewhat different ideas developed in \cite{Sohnius,DKT}
for the case of 4D, $\cN=2$ supersymmetric theories.

When $\D\neq 0$, there exists one more useful representation 
(in addition to the $\t$-frame and $\lambda$-frame) for the covariant 
derivatives 
\be
{\cal D}_{\bA} \quad \mapsto \quad 
{\bm \nabla}_{\bA}=
{\rm e}^{{\rm i}(\Omega +\S)}\,  \cD_{\bA}\,
{\rm  e}^{-{\rm i}(\Omega+ \S)}
\equiv \bD_{\bA} +{\rm i} \, \cV_{\bA}
~, \qquad \S = - \q^{- \hat \a} \q^+_{\hat \a} \,\D~.
\ee 
${}$For the operators 
$\bD_{\bA}= {\rm e}^{{\rm i}\S}\,  D_{\bA}\,
{\rm  e}^{-{\rm i}\S}$ one obtains
\bea 
{\bm \nabla}^+_{\hat \a} &=& 
\bD^+_{\hat \a} = \frac{\pa} {\pa \q^{- \hat \a} }~, \\
\bD^{++}&=& D^{++}+ {\rm i} \,(\hat{\q}^+)^2 \, \D~,
\quad
D^{++}=
u^{+i}\frac{\partial}{\partial u^{- i}} 
+{\rm i} \,(\G^{\hat a})_{\hat \b \hat \g} \,  
\q^{+\hat \b} \q^{+\hat \g} \,\frac{\pa}{\pa \bm{x}^{\hat a}}
+\q^{+\hat \a} \frac{\pa }{\pa \q^{- \hat \a} }~,
\non
\eea
where
\be
(\hat{\q}^+)^2 = \q^{+\hat \a} \q^+_{\hat \a}~.
\ee
As is seen, in this frame
the spinor gauge-covariant derivative ${\bm \nabla}^+_{\hat \a} $ coincides 
with partial derivatives with respect to $\q^{- \hat \a} $, while the 
analyticity-preserving gauge-covariant derivative  ${\bm \nabla}^{++}=
\bD^{++}+{\rm i}\,\cV^{++}$  acquires a term proportional to the central charge.

In accordance with \cite{DKT}, the supersymmetric action 
involves a special  gauge-invariant analytic superfield $\bL^{++} $
\be 
\bD^+_{\hat \a} \bL^{++}  =0 ~, 
\qquad \bD^{++}  \bL^{++}  =0 ~.
\ee
The action is
\be  
S=  {\rm i} \int {\rm d} \z^{(-4)} \,
(\hat{\q}^+)^2 \,\bL^{++}~.
\label{cc-action1}
\ee
Although $S$ involves naked Grassmann variables, it turns out to be 
supersymmetric,due to the constraints imposed on $ \bL^{++}$.
Its invariance under the supersymmetry transformations 
can be proved in complete analogy with 
the four-dimensional case \cite{DKT}.
The action  (\ref{cc-action1}) possesses another nice representation
obtained in Appendix C, eq. (\ref{RedAction2}).

One can transform $ \bL^{++} $ to the $\t$-frame in which 
 \be
L^{++} (z,u)={\rm e}^{-{\rm i}\S}\,  \bL^{++}
=L^{ij}(z)\,u^+_iu^+_j~.
\ee
This gauge-invariant superfield obeys the constrains
\be 
D^+_{\hat \a} L^{++}  =0 ~, 
\qquad D^{++}  L^{++}  =0 ~.
\label{lagrange-econstraint}
\ee
Doing the Grassmann and harmonic integrals in 
(\ref{cc-action1}) gives 
\be
S=\frac{\rm i}{12}  \int {\rm d}^5x \, \hat{\cD}^{ij} 
\,L_{ij} \double{\Big |}~, \qquad \quad
\hat{\cD}^{ij} = \cD^{\hat \a (i}
\cD^{j)}_{\hat \a} ~,
\label{cc-action2}
\ee
where we have replaced, for convenience,  
 ordinary spinor covariant derivatives 
by  gauge-covariant ones (this obviously does not change the action, 
for $L_{ij} $ is gauge invariant). 

In four space-time dimensions, the super-action (\ref{cc-action2}) 
was postulated by Sohnius \cite{Sohnius} several years before 
the discovery of harmonic superspace. 
It is quite remarkable that only within the harmonic superspace approach, 
this super-action can be represented as a superspace integral having a 
transparent physical interpretation. To wit, the factor 
${\rm i} (\hat{\q}^+)^2$ in (\ref{cc-action1}) 
can be identified with  a vacuum expectation value 
$\langle \cV^{++}_\D \rangle $ of the central-charge gauge superfield
$\cV^{++}_\D$ (compare with   (\ref{WZ-gauge})).
With such an interpretation, the super-action admits
simple generalizations to the cases of (i) rigid supersymmetric
theories with gauged central charge \cite{DIKST}, 
and (ii)  supergravity-matter systems \cite{KT}.

The super-action (\ref{cc-action1}), and its equivalent form (\ref{cc-action2}), 
can be used for superymmetric theories without central charge; an example
will be given below. It is only the constraints (\ref{lagrange-econstraint}) 
which are relevant in the above construction.

\subsection{Fayet-Sohnius hypermultiplet}
An example of a theory with non-vanishing central charge 
is provided by the Fayet-Sohnius hypermultiplet.
The Fayet-Sohnius hypermultiplet coupled to a 
Yang-Mills supermultiplet is described by the
Lagrangian  \cite{Sohnius,DKT}
\be
L^{++}_{\rm FS} = \hf \,
\breve{{\bm q}}^+ 
\stackrel{\longleftrightarrow}{ \D} 
{\bm q}^+ 
-{\rm i}\, m\, \breve{{\bm q}}^+ {\bm q}^+~,
\label{FS-Lagrangian}
\ee
with $m$ the hypermultiplet mass.

To compute the component action that follows from 
(\ref{FS-Lagrangian}), one should use the definitions 
(\ref{W-components}) and (\ref{FS-components}) 
for the component fields of $\cW$ and ${\bm q}^i$, 
respectively. 
\bea
S_{\rm FS} &=&\int {\rm d}^5x \, 
\Big\{ 
- \cD_a {\bar C}_k \cD^a C^k
-{\rm i}\bar \lambda 
{\not{\hbox{\kern-4pt $\cD$}}} 
\lambda
+ \bar F_k F^k + m\,\bar\lambda \lambda 
+{\bar \lambda}\vf \lambda
-{{\rm i}\over 2}\bar C_k\,X^k{}_\ell\, C^\ell
\non \\
&&\qquad 
-\hf \, {\bar C}_k \vf^2 C^k
 - \Big( {\rm i} \, m \,\bar F_k C^k 
-{1\over \sqrt{8} }\, \bar \lambda \J_k C^k
+{\rm i}\, \bar F_k \vf\, C^k
+{\rm c.c.} \Big)
\Big\} ~.
\eea

\subsection{Vector multiplet}
The Yang-Mills supermultiplet is described
by the Lagrangian\footnote{In terms of the
analytic prepotential $\cV^{++}$, 
the super Yang-Mills  action is non-polynomial
\cite{Z2}.}
\be
L^{++}_{\rm YM} = {1\over 4}\,
{\rm tr} \, \cG^{++} 
~, \qquad \D \, L^{++}_{\rm YM} =0~,
\label{SYM-action1}
\ee
with $\cG^{++} $ given in (\ref{YML}).
The corresponding equation of motion  can be shown 
to be 
\be
({\hat \cD}^+)^2  \cW =0
\quad \Leftrightarrow \quad 
\cD^+_{\hat \a} \cD_{\hat \b }^+   \cW
= 0~.
\label{on-shell-vm}
\ee
It follows from this that
\be
\cD^{\hat a} \cD_{\hat a}\, \cW = 
\hf \, \{ \cD^{\hat \a}_i \cW\, , \cD_{\hat \a}^i \cW \} ~.
\ee
In the Abelian case, eq. (\ref{on-shell-vm}) 
reduces to 
\be 
D^+_{\hat \a} D_{\hat \b }^+   \cW
= 0 \quad \Rightarrow \quad 
\pa^{\hat a} \pa_{\hat a}\, \cW = 0~.
\ee
${}$From the point of view of 4D, $\cN=2$ supersymmetry, 
this can be recognized as the off-shell superfield constraints
\cite{GHH,DKT,BHO}
describing the so-called linear vector-tensor multiplet
discovered by Sohnius, Stelle and West  
\cite{SSW} and re-vitalized fifteen years later
in the context of superstring compactifications
\cite{deWKLL}.

The Yang-Mills action with components defined by (\ref{WZ-V++}) is
\bea
S_{\rm YM}= \int {\rm d}^5x \, {\rm tr}\,
\Big\{-{1\over 4}\, F_{\hat a\hat b} F^{\hat a\hat b}
&-&{1\over 2} \cD_{\hat a} \vf \cD^{\hat a} \vf 
+{1\over 4}X^{ij}X_{ij} 
\non \\	
&+&
{{\rm i}\over 2} \J^k {\not{\hbox{\kern-4pt $\cD$}}}\J_k
	-{1\over 2} \J^k\left[ \vf, \J_k\right] 
\Big\} ~.
\eea

\sect{Chern-Simons Couplings}
\label{section:four}

Consider two 
vector multiplets:
(i) a U(1)  vector multiplet $\cV^{++}_\D$; 
(ii) a Yang-Mills vector multilpet $\cV^{++}_{\rm YM}$. 
They can be coupled to each other, in a gauge-invariant 
way, using the interaction 
\be
S_{\rm int} = \int {\rm d} \z^{(-4)} \, \cV^{++}_\D \, 
{\rm tr} \, \cG^{++}_{\rm YM} ~,
\label{CS-inter}
\ee
where  $\cG^{++}_{\rm YM} $ corresponds
to the non-Abelian multiplet and  is defined 
as in eq. (\ref{YML}).  Invariance of $S_{\rm int} $ 
under the U(1) gauge transformations 
\be
\d \cV^{++}  =- D^{++} \lambda~, \qquad D^+_{\hat \a} \lambda =0~,
\ee
follows from the constraints (\ref{YML}) to which
$\cG^{++}_{\rm YM} $ is subject.

Let us assume that  the physical scalar field in 
$\cV^{++}_\D$ possesses a non-vanishing expectation value 
(such a situation occurs, for instance, 
when $\cV^{++}_\D$ is the vector multiplet 
gauging the central charge symmetry). 
In accordance with \cite{vev}, 
this condition is expressed as 
$\langle \cW_\D (z) \rangle =\m \neq 0$; 
then, there exists a gauge fixing such that 
\be
\langle  \cV^{++}_\D (\z, u) \rangle 
={\rm  i}\, \m \,(\hat{\q}^+)^2 ~, \qquad 
\m ={\rm const}~.
\ee
Now, combining the interaction (\ref{CS-inter}) 
with the gauge-invariant kinetic terms for $\cV^{++}_\D$ 
and $\cV^{++}_{\rm YM}$, the complete action becomes
\be
S =  \int {\rm d} \z^{(-4)} \, \cV^{++}_\D \,
\Big\{  g^{-2}_\D\,
\cG^{++}_\D + g^{-2}_{\rm YM} \,
{\rm tr} \, \cG^{++}_{\rm YM} \Big\}~,
\label{CS}
\ee
with $g_\D$ and $g_{\rm YM} $ coupling constants.
A different form for this action was given in \cite{Z}.
The theory (\ref{CS}) is superconformal at the classical level. 
It would be interesting to compute, for instance, 
perturbative quantum corrections.

Let us consider the special case of a single Abelian gauge 
field, 
$ \cV^{++}_\D= \cV^{++}_{\rm YM} \equiv \cV^{++}$.
The equations of motion for the corresponding 
Chern-Simons theory, 
\be
S_{\rm CS} =  {1\over 12 g^2}
\int {\rm d} \z^{(-4)} \, \cV^{++} \,
\cG^{++} ~,
\label{CS2}
\ee
can be shown to be 
\be 
-{\rm i} \, \cG^{++} = 
\cD^{+ \hat \a} \cW \, \cD^+_{\hat \a} \cW 
+\hf \,\cW \,  ({\hat \cD}^+)^2 \cW =0 ~. 
\ee
Using the Bianchi identity (\ref{Bianchi2}), 
one can rewrite this in the form 
\be 
D^+_{\hat \a} \,D^+_{\hat \b} \,\cW 
= -\hf \,\ve_{\hat \a \hat \b} \,
\frac{ D^{+ \hat \g} \cW \, D^+_{\hat \g} \cW }
{\cW}~.
\label{nvt}
\ee
${}$From the point of view of 4D, $\cN=2$ supersymmetry, 
this can be recognized as the off-shell superfield constraint
describing the so-called 
nonlinear vector-tensor multiplet\footnote{The
nonlinear vector-tensor multiplet was discovered
in \cite{CdWFKST}.}
\cite{DK+IS,DIKST}.
Resorting to  the two-component spinor notation, 
eq.  (\ref{nvt}) leads to 
\bea
D^+_\a {\bar D}^+_{\dt \a} \cW =0~, 
\qquad 
D^{+\a} D^+_\a \cW
= -{1\over \cW} \Big( D^{+ \a} \cW \,D^+_\a \cW
-{\bar D}^+_{\dt \a} \cW \,{\bar D}^{+ {\dt \a}} \cW \Big)~.
\eea

In the case of the dynamical system (\ref{CS}), 
the equation of motion for the Abelian gauge field is 
\be
{1\over \k}\, \cG^{++}_\D = 
{\rm tr} \, \cG^{++}_{\rm YM} ~,
\ee 
with $\k$ a coupling constant. With properly 
defined dimensional reduction 5D $\to$ 4D, 
this can be recognized  as the superfield constraint describing 
the Chern-Simons coupling of a nonlinear vector-tensor
to an external  $\cN=2$ Yang-Mills supermultiplet \cite{DIKST}.

The super Chern-Simons actions  can be readily 
reduced to components in the Wess-Zumino gauge  
(\ref{WZ-V++}) for the Abelian gauge field $\cV^{++}$.
If $L^{++}(z,u)= L^{ij}(z)\,u^+_iu^+_j$ is a real analytic
superfield of the type (\ref{lagrange-econstraint}), then  
\bea
S&=& \int {\rm d}\z^{(-4)} \,\cV^{++} L^{++} \\
&=& \int {\rm d}^5 x \,
\Big\{ X^{ij} L_{ij}
+{ {\rm i} \over 12} \vf\, \hat \cD^{ij} L_{ij}
+{{\rm i}\over 12} \,A_{\hat a}\, (\cD^i \G^{\hat a} \cD^j) L_{ij}
-{2\over 3} \,\J^{i \hat \a}\, \cD^j_{\hat \a} L_{ij} 
\Big\} \double{\Big|}~.
\non 
\eea	
The Abelian supersymmetric Chern-Simons theory
(\ref{CS2}) leads to the following component action:
\bea
S_{\rm CS}
&=& {1\over 2 g^2} \int {\rm d}^5 x \left\{
{1\over 3} \, \e^{\hat a\hat b\hat c\hat d\hat e}\,
A_{\hat a}F_{\hat b\hat c}  F_{\hat d \hat e}
-{1\over 2} \,\vf F_{\hat a \hat b}F^{\hat a \hat b} 
-\vf \pa_{\hat a}\vf \pa^{\hat a} \vf 
+{1\over 2}\, \vf X_{ij} X^{ij}
\right. \non \\
&&	\qquad \qquad \left.
-{{\rm i}\over 2}\, F_{\hat a \hat b}(\J^k\S^{\hat a \hat b} \J_k)
+{\rm i}\, \vf(\J^k  {\not{\hbox{\kern-2pt $\partial$}}} \J_k)
-{{\rm i}\over 2} \,X_{ij} (\J^i \J^j)\right\} ~.
\eea

\sect{5D Supermultiplets in Reduced Superspace}
\label{section:five}

Some of the results described in the previous sections
can easily be reduced to a ``hybrid'' formulation 
which keeps manifest only 4D, $\cN=1$ super 
Poincar\'e symmetry. As the 5D superfields depend
on two sets of 4D anticommuting Majorana
spinors, $(\q^\a_{\1}\,, {\bar \q}^{\1}_{\dt \a})$
and $(\q^\a_{\2}\,, {\bar \q}^{\2}_{\dt \a})$, 
such a hybrid formulation is equivalent to 
integrating out, say,  the second set
and keeping intact the first set of variables
\be
\q^\a = \q^\a_{\1} ~, \qquad {\bar \q}_{\dt \a}=
{\bar \q}_{\dt \a}^{\1}~.
\label{theta1}
\ee 
In this approach, one deals  with 
reduced  (or $\cN=1$) superfields
$U | $, $ D^{\2}_\a U|$, $ {\bar D}_{\2}^{\dt \a} U|, \dots$
(of which not all are usually independent)    
and 4D, $\cN=1$ spinor covariant derivatives $D_\a$ 
and ${\bar D}^{\dt \a}$ defined in the obvious way:
\be 
U| = U(x, \q^\a_i, {\bar \q}^i_{\dt \a})
\Big|_{ \q_{\2} = {\bar \q}^{\2}=0 }~,
\qquad D_\a = D^{\1}_\a \Big|_{\q_{\2} ={\bar \q}^{\2}=0} ~, 
\qquad
{\bar D}^{\dt \a} = {\bar D}_{\1}^{\dt \a} 
\Big|_{\q_{\2} ={\bar \q}^{\2}=0}~.
\label{N=1proj}
\ee 
Our consideration below naturally reproduces many
of the  5D supersymmetric models 
originally derived in the hybrid formulation \cite{hybrid}.

\subsection{Vector multiplet}
\label{section:SimpleVector}

Let us introduce reduced gauge covariant derivatives 
\be
\Big\{ {\bm \cD}_{\a}, {\bar {\bm{\cD}}}^{\dt \a}, {\bm \cD}_{a}, {\bm \cD}_5 \Big\}
~=~
\Big\{ \cD^{\1}_{\a}, \cD_{\1}^{ \dt \a}, \cD_{a}, \cD_5 \Big\} \Big|~.
\ee
As follows from  (\ref{SYM-algebra}), their algebra is
\bea
& \{ {\bm \cD}_\a \,, {\bm \cD}_\b \} 
= \{ {\bar {\bm \cD}}_{\dt \a} , {\bar {\bm \cD}}_{\dt \b} \} =0~, \qquad 
\{ {\bm \cD}_\a \,, {\bar {\bm \cD}}_{\dt \b} \} = - 2{\rm i} \, {\bm \cD}_{\a {\dt \b}}~, \non \\
& [ {\bm \cD}_\a \,, {\bm \cD}_{\b {\dt \b}}] = -2 {\rm i} \ve_{\a \b}\,{\bar \cW}_{\dt \b} ~, 
\qquad 
[{\bar {\bm \cD}}_{\dt \a} \,, {\bm \cD}_{\b {\dt \b}}] = 
-2{\rm i} \ve_{{\dt \a} {\dt \b}}\,\cW_\b ~ , 
\non \\
& [ {\bm \cD}_\a \,, {\bm \cD}_5 + \cF] = 0~,
\qquad 
 [ {\bar {\bm \cD}}_{\dt \a} \,, {\bm \cD}_5 -\cF] =  0~,
\label{N=1cov-der-al}
\eea
where
\bea
\cF=\cW |~, \qquad  \cW_\a=\cD^{\2}_\a \cW|~.
\label{N=1field-strengths}
\eea
It can be seen that the field strengths 
$\cF$, $\cW_\a$ and ${\bar \cW}^{\dt \a}$
are the only independent $\cN=1$ descendants of $\cW$.

The strengths $\cF$ and $\cW_\a$ obey some constraints which 
follow from the Bianchi identities
(\ref{Bianchi1}) and (\ref{Bianchi1.5}).
Consider first the constraint 
(\ref{Bianchi1}) with two derivatives of $\cW$. 
Taking the part with $(i,j ,\hat \a, \hat \b)=(\1, \1, \a, {\dt \a})$ 
gives the ``$\cN=1$ chirality'' of $\cW_\a$
\bea
\bar {\bm \cD}_{\dt \a} \cW_\a=0~.
\label{BIa}
\eea
Taking instead the part with $(i,j ,\hat \a, \hat \b)=(\1, \2, \a, \b)$ gives the familiar 
Bianchi identity
\bea
{\bm \cD}^\a \cW_\a-\bar {\bm \cD}_{\dt \a} \bar \cW^{\dt \a}=0~.
\label{BIb}
\eea
${}$Next, the $(i,j ,\hat \a, \hat \b)=(\1, \1, \a, \b)$ and 
$(i,j ,\hat \a, \hat \b)=(\1, \2, \a, {\dt \a})$ parts, 
respectively, give 
\bea
\bar  \cD_{\2 \dt \g}\bar \cD_{\2}^{ \dt \g} \cW|=-{\bm \cD}^2 \cF~,
 \qquad 
\cD^{\2}_\a \bar \cD_{\2 \dt \b} \cW|={\bm \cD}_\a 
\bar {\bm \cD}_{\dt \a} \cF~.
\eea
The latter  identities support the  statement that 
$\cF$, $\cW_\a$ and ${\bar \cW}^{\dt \a}$
are the only independent $\cN=1$ descendants of $\cW$.
${}$Finally, decomposing the second constraint 
(\ref{Bianchi1.5}) with $(i,j,k)=(\2,\2,\1)$ and $(\hat \a, \hat \b, \hat \g)
=(\dot \a ,\dot \b,\g)$ gives 
\bea
-{1\over 4} \bar {\bm \cD}^2{\bm \cD}_\a \cF
+{\bm \cD}_5 \cW_\a-\left[\cF , \cW_\a\right]=0~.
\label{BIc}
\eea

In accordance with (\ref{SYM-action1}),
the super Yang-Mills action is 
\be
S_{\rm YM} = \frac{\rm i}{12}  \int {\rm d}^5x \, \hat{\cD}_{ij} 
L^{ij}_{\rm YM} \double{\Big|}~, \qquad 
L^{ij}_{\rm YM}= 
\frac{\rm i}{4} \, {\rm tr} \Big(\cD^{{\hat \a}i} \cW \, \cD^j_{\hat \a} \cW 
+{1 \over 4} \{\cW \,, 
{\hat \cD}^{ij} \cW \}\Big)~.
\ee
Its reduced form can be shown to be
\bea
S_{\rm YM}&=&{\rm tr} \int {\rm d}^5 x \, \left\{ 
{1\over 4}\int {\rm d}^2\q \, \cW^\a \cW_\a
+{1\over 4}\int {\rm d}^2 \bar \q \,\bar \cW_{\dt \a} \bar \cW^{\dt \a} 
+\int {\rm d}^4\q \,\cF^2
\right\}~.
\label{eqn:SimpleYM}
\eea
Here the Grassmann measures
${\rm d}^2\q$ and ${\rm d}^4\q$ are part of 
the chiral and the full superspace measures, respectively,  
in 4D, $\cN=1$ supersymmetric field theory.

It is instructive to solve the constraints encoded in (\ref{N=1cov-der-al}).
A general solution to the equations $\{ {\bm \cD}_\a \,, {\bm \cD}_\b \} =
 [ {\bm \cD}_\a \,, {\bm \cD}_5 + \cF] = 0$ is
 \be
{\bm \cD}_\a =  {\rm e}^{-\X} \,D_\a \, {\rm e}^{\X}~,
\qquad 
{\bm \cD}_5 + \cF = {\rm e}^{-\X} \,\Big(\pa_5  +\F^\dagger \Big)\, {\rm e}^{\X}
~, 
\qquad 
D_\a \F^\dagger =0~,
\ee
for some Lie-algebra-valued prepotentials $\X$ and $\F^\dagger$, 
of which $\X$  is complex unconstrained and $\F^\dagger $ 
antichiral.
Similarly, the constraints 
$ \{ {\bar {\bm \cD}}_{\dt \a} , {\bar {\bm \cD}}_{\dt \b} \} =
 [ {\bar {\bm \cD}}_{\dt \a} \,, {\bm \cD}_5 -\cF] =  0$ 
 are solved by 
 \be
{\bar {\bm \cD}}_{\dt \a} =  {\rm e}^{\X^\dagger} \,{\bar D}_{\dt \a} \, {\rm e}^{-\X^\dagger}~,
\qquad 
{\bm \cD}_5 -\cF =  {\rm e}^{\X^\dagger} \,\Big(\pa_5  -\F \Big)
 \, {\rm e}^{-\X^\dagger}~, \qquad {\bar D}_{\dt \a} \F=0~.
\ee
The prepotentials introduced possess the following 
gauge transformations
\bea
 {\rm e}^{\X^\dagger} ~&\to & ~ {\rm e}^{{\rm i} \t (z)}\,
 {\rm e}^{\X^\dagger}\,{\rm e}^{-{\rm i} \l (z)}~,
 \quad 
\F ~\to  ~ 
 {\rm e}^{{\rm i} \l (z)}\Big(\F- \pa_5 \Big)\,{\rm e}^{-{\rm i} \l (z)} 
\cdot 1
~, 
\quad {\bar D}_{\dt \a} \l =0~.
\eea
Here the $\l$-gauge group occurs as a result of solving 
the constraints in terms of the unconstrained prepotentials.

By analogy with the 4D $\cN=1$ super Yang-Mills case,
one can introduce a chiral representation defined by
applying a {\it complex} gauge transformation with
$\t = -\X^\dagger$.
This gives
\bea
{\bm \cD}_\a =  {\rm e}^{-V} \,D_\a \, {\rm e}^{V}~,
\quad && \quad
{\bar {\bm \cD}}_{\dt \a} =  {\bar D}_{\dt \a} ~,\non \\
{\bm \cD}_5 + \cF = {\rm e}^{-V} \,\Big(\pa_5  +\F^\dagger \Big)\, {\rm e}^{V}
\quad && \quad
{\bm \cD}_5 -\cF =  
\pa_5  -\F ~,
\eea
where
\bea
{\rm e}^{V}= {\rm e}^{\O}\,{\rm e}^{\O^\dagger}~,
\qquad V^\dagger =V~.
\non
\eea
Here the real Lie-algebra valued superfield $V$ is
the standard $\cN=1$ super Yang-Mills prepotential. For $\cF$ 
we obtain
\be
2 \,\cF= 
 \F +   {\rm e}^{-V} \,\F^\dagger \, {\rm e}^{V} 
+ {\rm e}^{-V} \,(\pa_5 \, {\rm e}^{V} )~.
\ee
We have thus reproduced the results obtained by Hebecker 
within the hybrid approach \cite{hybrid}. 

\subsection{Fayet-Sohnius hypermultiplet}

The Fayet-Sohnius hypermultiplet ${\bm q}^i$ generates two 
independent
$\cN=1$ superfields transforming in the same representation 
of the gauge group,
\bea 
\tilde{Q}^\dagger &=& {\bm q}^{\1} |~,
\qquad 
{Q} = {\bm q}^{\2} |~,
\eea
and obeying the constraints 
\be
 {\bm \cD}_\a \tilde{Q}^\dagger =0~, 
 \qquad {\bar {\bm \cD}}_{\dt \a} Q=0~.
 \ee
 These constraints follow from (\ref{FSh}).
 Thus $Q$ and $\tilde{Q}^\dagger $ are covariantly 
chiral and antichiral,  respectively.
 The central charge transformation of these superfields is:
 \bea
{\rm i} \,\D Q &=&{1\over 4}\bar {\bm \cD}^2 \tilde{Q}^\dagger 
+ (\cF - {\bm \cD}_5) Q~, \qquad
{\rm i} \,  \D \tilde{Q}^\dagger
={1\over 4} {\bm \cD}^2 Q +
(\cF + {\bm \cD}_5) \tilde{Q}^\dagger ~.
\eea

In accordance with (\ref{FS-Lagrangian}),
the action for the Fayet-Sohnius hypermultiplet is
\bea
S_{\rm FS}&=&{{\rm i}\over 12} \int {\rm d}^5 x \, 
\hat{\cD}_{ij} L^{ij}_{\rm FS}\Big|~,\qquad 
L^{ij}_{\rm FS}=-\Big(
{1\over 2}\bar {\bm q}^{(i} \stackrel{\leftrightarrow}{\D} {\bm q}^{j)}
-{\rm i}\, m \bar {\bm q}^{(i} {\bm q}^{j)}\Big)~.
\eea 
It  can be shown to reduce to the following $\cN=1$ action 
\bea
S_{\rm FS}&=&
\int {\rm d}^5 x \left\{ \int {\rm d}^4 \q \,( 
Q^\dagger \, Q + \tilde{Q} \,  \tilde{Q}^\dagger )
+ \Big( \int {\rm d}^2 \q\,  	
\tilde{Q} (\cF -{\bm \cD}_5 +m) Q
+{\rm c.c.}\Big) \right\} ~.
\label{FS-actionN=1}
\eea
As follows from (\ref{N=1cov-der-al}), the operator 
${\bm \cD}_5 -\cF$ preserves chirality.

\sect{Projective Superspace and Dimensional Reduction}
\label{section:six}

Throughout this section, 
we consider only 5D supermultiplets 
without central charge, $\D = 0$.
However, many results below can be extended to include 
the case $\D\neq 0$.

\subsection{Doubly punctured harmonic superspace} 

Let ${\bm \j}^{(p)} (z,u)$ be a harmonic superfield
of non-negative U(1) charge $p$. 
Here we will be interested 
in solutions to the equation
\be
D^{++} \, {\bm \j}^{(p)}= 0\quad \Rightarrow \quad
D^{++} D^+_{\hat \a} \,{\bm \j}^{(p)}= 0~, \qquad p\geq 0~.
\label{holom1}
\ee

If ${\bm \j}^{(p)} (z,u)$ is 
globally defined and smooth
over ${\mathbb R}^{5|8} \times  S^2$, 
it possesses a convergent Fourier series of the form 
(\ref{smoothfunction}). 
If ${\bm \j}^{(p)} (z,u)$ is further 
constrained  to obey the
equation (\ref{holom1}), 
then its general form  becomes
\be
{\bm \j}^{(p)}(z,u) = 
{\bm \j}^{i_1 \cdots i_{p} } (z)\,
u^+_{i_1} \cdots u^+_{i_{p}}~.
\ee
Therefore, such a globally defined harmonic superfield 
possesses finitely many component fields, and this can 
thought of as a consequence of the Riemann-Roch 
theorem \cite{FK} specified to the case of  $ S^2$.
A more interesting situation occurs if one allows
 ${\bm \j}^{(p)} (z,u)$ to have a few singularities on $ S^2$.

${}$For further consideration, 
it is useful to cover $ S^2$ by two charts 
and introduce local complex coordinates in each chart, 
as defined in  Appendix B. In the north chart (parametrized 
by the complex variable $w$ and its conjugate $\bar w$) 
we can represent ${\bm \j}^{(p)} (z,u)$ as follows
\be
{\bm \j}^{(p)}(z,u) = (u^{+\1})^p \, \j(z,w, {\bar w})~.
\ee
If ${\bm \j}^{(p)} (z,u)$ is globally defined over
${\mathbb R}^{5|8} \times  S^2$, 
then $\j(z,w, {\bar w}) \equiv  
{\bm \j}^{(p)}_{\rm N}(z,w, {\bar w})$
is given as in eq. (\ref{fn}).
It is a simple exercise to check that
\be
D^{++} {\bm \j}^{(p)} (z,u) = (u^{+\1})^{p+2}\, (1+ w {\bar w})^2  \,
\pa_{\bar w}\, \j (z,w, {\bar w})~, 
\qquad p\geq 0~,
\label{dplus}
\ee
and therefore
\bea
D^{++} \, {\bm \j}^{(p)} = 0~, \quad p\geq0 
\quad  \Leftrightarrow \quad 
\pa_{\bar w} \,\j  =0~.
\eea
Assuming that ${\bm \j}^{(p)} (z,u)$ may possess 
singularities only at the north and south poles 
of $ S ^2$, we then conclude that
\be
\j (z, w) = \sum_{n=-\infty}^{+\infty} \j_n (z) \,w^n~.
\ee

Now, consider an analytic superfield ${\bm \f}^{(p)}$ 
obeying the constraint (\ref{holom1}). 
\be
D^+_{\hat \a} \, {\bm \f}^{(p)}= 0~, \qquad 
D^{++} \, {\bm \f}^{(p)}= 0~, 
\qquad  p \geq 0~.
\label{holom2}
\ee
We assume that ${\bm \f}^{(p)} (z,u)$  
is non-singular outside the north and south poles 
of $ S ^2$. Then, representing
${\bm \f}^{(p)}(z,u) = (u^{+\1})^p \, \f(z,w, {\bar w})$ and defining 
\be
D^+_{\hat \a} = - u^{+\1}\,  \nabla_{\hat \a} (w)~, 
\qquad \nabla_{\hat \a} (w) = -D^i_{\hat \a} \, w_i~,
\qquad w_i = (-w, 1)~,
\ee
eq. (\ref{holom2}) is solved as 
\be
\nabla_{\hat \a} (w) \, \f(z,w)=0~, \qquad
\f (z, w) = \sum_{n=-\infty}^{+\infty} \f_n (z) \,w^n~.
\label{holom0}
\ee
These relations define a {\it projective superfield}, following 
the four-dimensional terminology \cite{projective}.
Since the supersymmetry transformations 
act simply as the identity transformation on $ S ^2$, 
the above consideration clearly defines 
supermultiplets. Such supermultiplets turn out 
to be most suited for dimensional reduction.

The projective analogue of the smile-conjugation 
(\ref{breve}) is 
\bea 
\breve{\f} (z, w) = \sum_{n=-\infty}^{+\infty} (-1)^n \,
{\bar  \f}_{-n} (z) \,w^n~, \qquad 
\nabla_{\hat \a} (w) \, \breve{\f}(z,w)=0~.
\label{holom3}
\eea
If $\breve{\f} (z, w) = {\f} (z, w) $, the projective superfield 
is called real. The projective conjugation (\ref{holom3})
can be derived from the smile-conjugation 
(\ref{breve}), see \cite{Kuzenko:1998xm}
for details.

There are several types of projective superfields \cite{projective}. 
A real projective superfield of the form (\ref{tropical})
is called a {\it tropical}  multiplet. A real  projective superfield of the form 
\be
\f (z, w) = \sum_{-n}^{+n} \f_n (z) \,w^n~, \qquad 
\breve{\f}  = {\f} ~,\qquad \qquad n \in {\mathbb Z}
\label{holom4}
\ee
is called a real O($2n$) multiplet.\footnote{One can also 
introduce complex O($2n$+1) multiplets \cite{projective}.} 
A projective superfield $\U(z,w)$
of the form (\ref{pm}) is called an {\it arctic} multiplet, 
and its conjugate,  $\breve{\U}(z,w)$, an {\it antarctic} multiplet.
The  $\U(z,w)$ and $\breve{\U}(z,w)$ constitute a {\it polar} multiplet.
More general projective superfields occur if one multiplies any 
of the considered superfields by $w^n$, with $n$ an integer.

At this stage, it is useful to break the manifest 
5D Lorentz invariance by 
switching  from the four-component spinor notation 
to the two-component one. Representing
\bea
\nabla_{\hat \a} (w) 
= \left(
\begin{array}{c}
\nabla_\a (w) \\
{\bar \nabla}^{\dt \a}  (w)  
\end{array}
\right)~, 
\quad \nabla_\a (w) \equiv  w D^{\1}_\a - D^{\2}_\a ~,
\quad
{\bar \nabla}^{\dt \a} (w) \equiv {\bar D}^{\dt \a}_{ \1} + 
w {\bar D}^{\dt \a}_{ \2}~,
\label{nabla}
\eea
the constraints (\ref{holom3})
can be rewritten in the component form
\be
D^{\2}_\a \f_n = D^{\1}_\a \f_{n-1} ~,\qquad
{\bar D}^{\dt \a}_{\2} \f_n = - {\bar D}^{\dt \a}_{ \1} 
\f_{n+1}~.
\label{pc2}
\ee
In accordance with (\ref{4D-N2covder1}) and
(\ref{4D-N2covder2}), 
one can think of the operators
$D_A = (\pa_a , D^i_\a , {\bar D}^{\dt \a}_i )$, 
where $a=0,1,2,3$, 
as the  covariant derivatives
of  4D, $\cN=2$ central charge superspace, 
with $x^5$ being the central charge variable.
The relations (\ref{pc2}) imply  that the dependence
of the component superfields 
$\f_n$ on $\q^\a_{\2}$ and ${\bar \q}^{\2}_{\dt \a}$ 
is uniquely determined in terms 
of their dependence on $\q^\a_{\1}$
and ${\bar \q}^{\1}_{\dt \a}$.  In other words, 
the projective superfields depend effectively 
on half the Grassmann variables which can be choosen
to be the spinor  coordinates of 4D, $\cN=1$ superspace
(\ref{theta1}).
In other words, it is sufficient to work with reduced  superfields
$\f(w) | $ and 4D, $\cN=1$ spinor covariant derivatives $D_\a$ 
and ${\bar D}^{\dt \a}$ defined in (\ref{N=1proj}).

If the series  in  (\ref{holom0})
 is bounded from below (above), then eq. (\ref{pc2}) implies that
 the two 
 lowest (highest) components in $\f(w) | $ are constrained 
 $\cN=1$ superfields.  For example,  in the case of the 
arctic multiplet, eq.  (\ref{pm}), 
the leading components 
$\F = \U_0 | $ and $\G = \U_1 |$ 
obey the constraints (\ref{pm-constraints}).

Given a real projective superfield $L (z,w)$, 
one can construct a supersymmetric invariant
\be 
S = \frac{1}{2\pi \,{\rm i} } \oint_{C} \frac{{\rm d}w}{w}\,
 \int {\rm d}^5 x \, {\rm d}^4 \q \,  L(w)\Big|
 \equiv  \frac{1}{2\pi \,{\rm i} } \oint_{C} \frac{{\rm d}w}{w}\, S(w)~,
\label{integral-projective}
 \ee
 with $C$ a contour around the origin
(in what follows, such a contour is always assumed). 
${}$For  $S(w) $ 
there are
several equivalent  forms:
\bea
S(w)= {1\over 16}  \int {\rm d}^5 x \, D^2 {\bar D}^2 \,
L (z,w)\double{\Big|}
= {1\over 16}  \int {\rm d}^5 x \, (D^{\1})^2 ({\bar D}_{\1})^2 \,L (z,w)\double{\Big|}
\label{integral-projective2}
\eea
assuming only that total space-time derivatives do not contribute.
The invariance of $S(w)$ under arbitrary SUSY transformations
is easy to  demonstrate. Defining 
\be
D^4 = {1\over16}  (D^{\1})^2 ({\bar D}_{\1})^2~,
\label{D4}
\ee
one can argue as follows:
\bea
\d S(w)  &=& {\rm i} \int {\rm d}^5 x\, \left(\ve^\a_i Q^i_\a +
{\bar \ve}^i_{\dt \a} {\bar Q}^{\dt \a}_i \right) D^4 L (z,w) \double{\Big|} 
= - \int {\rm d}^5 x\, \left(\ve^\a_i D^i_\a +
{\bar \ve}^i_{\dt \a} {\bar D}^{\dt \a}_i \right) D^4 L(z,w)\double{\Big|}
 \non \\
&=& - \int {\rm d}^5 x\, \left(\ve^\a_{\2} D^{\2}_\a +
{\bar \ve}^{\2}_{\dt \a} {\bar D}^{\dt \a}_{\2} \right) D^4 L(z,w) 
\double{\Big|} 
= - \int {\rm d}^5 x\, D^4 \left(\ve^\a_{\2} D^{\2}_\a +
{\bar \ve}^{\2}_{\dt \a} {\bar D}^{\dt \a}_{\2} \right) L(z,w)\double{\Big|}   
\non \\
&=& - \int {\rm d}^5x \, D^4 \left( \ve^\a_2 D^{\1}_\a \,w  
- {\bar \ve}^{\2}_{\dt \a} {\bar D}^{\dt \a}_{\1} \,w^{-1}  \right) 
L(z,w) \double{\Big|}= 0~,
\eea
with $Q_\a^i$ and ${\bar Q}_i^{\dt \a}$ the supersymmetry generators.

\subsection{Tensor multiplet and nonlinear sigma-models}

The tensor multiplet (also called O(2) multiplet) 
is described by a constrained
real analytic superfield $\Xi^{++}$:
\be
D^+_{\hat \a} \,\Xi^{++} =0~, \qquad D^{++} \,\Xi^{++} =0~.
\ee
The corresponding projective superfield $\X(z,w)$ 
is defined by 
$\Xi^{++}(z,u) = {\rm i}\, u^{+\1}u^{+\2}\, \X(z,w)$. 
Without distinguishing between $\X(z,w)$ and  $\X(z,w)|$,
we have 
\be
\X (w) ~=~ \F + w \,G - w^2 \,{\bar \F} ~,
\qquad {\bar G} =G~,
  \label{O2M}
\ee
where the component superfields obey the constraints
\be 
{\bar D}^{\dt \a} \, \F =0~,
\qquad
-{1\over 4} {\bar D}^2 \, G 
= \pa_5\,  \F~. 
\ee 

Here we consider a 5D generalization of the 4D, $\cN=2$ 
supersymmetric nonlinear sigma-model\footnote{The construction
given in  \cite{GHK} has recently been reviewed 
and extended in \cite{RVV}.} 
studied in \cite{GHK} 
and related  to the so-called $c$-map \cite{CFG}.
Let $F$ be a holomorphic function of $n $ variables. 
Associated with this function is the following supersymmetric action 
\be
S ~=~ - \int {\rm d}^5 x \, {\rm d}^4 \q \,  \Big[ \, 
\frac{1}{2\pi \,{\rm i} } \oint \frac{{\rm d}w}{w}
\, \frac{  F(\X^I (w))  }{w^2} ~+~ {\rm c.c.} \, \Big]~.
\label{tensoraction}
\ee
Since 
\bea
F(\X^I (w)) \,&=&\, F \Big( \F^I + w \,G^I - w^2 \,{\bar \F}^I \Big)
\non \\
&=&\, F(\F) + w \,F_I (\F) G^I - w^2 \,\Big( F_I (\F) {\bar \F}^I
- \hf \,F_{IJ} (\F) \, G^I G^J \Big)
~+~ O(w^3) ~, \non
\eea
the contour integral is trivial to do.  
The action is equivalent to
\be
S[\F, \bar \F, G]= \int {\rm d}^5 x \, {\rm d}^4 \q \, 
 \left\{ K (\F, {\bar \F})
- \hf \, g_{I {\bar J} } (\F, \bar \F ) \, G^I G^J \right\} ~,
\label{N2TMact} 
 \ee
where
\be
K(\F , \bar \F ) = {\bar \F}^I \, F_I (\F) + \F^I \,
{\bar F}_I (\bar \F) ~, \qquad g_{I {\bar J} } (\F, \bar \F )
=  F_{IJ}(\F) + {\bar F}_{IJ}(\bar \F) ~.
\label{c-kahler}
\ee
The K\"ahler potential $K(\F , \bar \F ) $ 
generates the so-called
rigid special K\"ahler geometry \cite{rsg}. 

Let us work out a dual formulation for the theory (\ref{N2TMact}). 
Introduce a first-order action
\bea 
&&S[\F, \bar \F, G] + \int {\rm d}^5 x \,\Big\{ 
\int {\rm d}^2 \q\, \J_I \Big( \pa_5\,  \F^I +
{1\over 4} {\bar D}^2 \, G^I \Big)
~+~{\rm c.c.} \Big\} \non \\
&=& S [\F, \bar \F, G] 
- \int {\rm d}^5 x \,\Big\{ 
\int {\rm d}^4 \q \, (\J_I  +{\bar \J}_I)\,G^I
+\Big( \int {\rm d}^2 \q\, \J_I \pa_5\,  \F^I 
\,+\,{\rm c.c.}\Big)  \Big\} ~,
\eea
where the superfield $G^I$ is now  real unconstrained, 
while $\J_I$ is chiral, ${\bar D}_{\dt \a} \J_I =0$.
In this action we can integrate out $G^I$ using the corresponding equations
of motion. This gives
\bea
S [\F, \bar \F, \J ,\bar \J ] =  \int {\rm d}^5 x \,\Big\{ 
\int {\rm d}^4 \q \,  H(\F, \bar \F, \J ,\bar \J ) 
+ \Big( \int {\rm d}^2 \q\, \J_I \pa_5\,  \F^I
\,+\,{\rm c.c.}\Big)  \Big\} ~,
\eea
where 
\be
H(\F, \bar \F, \J, \bar \J) = K(\F, \bar \F) + \hf\, g^{I {\bar
J} } (\F, \bar \F) ( \J_I + {\bar \J}_I) ( \J_J + {\bar \J}_J) ~.
\label{cmap}
\ee
The potential $H(\F, \bar \F, \J, \bar \J)$  is the K\"ahler potential  
of a hyper K\"ahler manifold. By construction, this potential is generated 
by another K\"ahler potential, $K(\F, \bar \F)$, which  is associated
with the holomorphic function  $F(\F)$ 
defining the rigid special K\"ahler geometry \cite{rsg}. 
The correspondence $K(\F, \bar \F) ~\to ~
H(\F, \bar \F, \J, \bar \J)$ is called the rigid $c$-map  \cite{CFG}.

\subsection{Polar hypermultiplet and nonlinear sigma-models}

According to \cite{projective}, the polar hypermultiplet is 
generated by projective superfields
\be
\U (z,w) = \sum_{n=0}^{\infty} \U_n (z) \,w^n~, \qquad
\breve{\U} (z,w) = \sum_{n=0}^{\infty} (-1)^n {\bar \U}_n (z)\,
\frac{1}{w^n} ~.
\label{pm}
\ee
The projective superfields $\U$ and $\breve{\U}$ are called arctic
and antarctic \cite{projective}, respectively.
The constraints (\ref{pc2}) imply that the leading components 
$\F = \U_0 | $ and $\G = \U_1 |$ are constrained 
\be 
{\bar D}^{\dt \a} \, \F =0~,
\qquad
-{1\over 4} {\bar D}^2 \, \G 
= \pa_5\,  \F~. 
\label{pm-constraints}
\ee 
The other components of $\U (w)$ are complex unconstrained superfields, 
and they appear to be non-dynamical (auxiliary)
 in models with at most two space-time derivatives
at the component level.

Here we consider a 5D generalization of 
the 4D, $\cN=2$ supersymmetric 
nonlinear sigma-model studied in \cite{GK}. 
It is described by the action 
\be
S[\U, \breve{\U}]  =  \int {\rm d}^5 x \, {\rm d}^4 \q \,  \Big[ \, 
\frac{1}{2\pi {\rm i}} \, \oint \frac{{\rm d}w}{w} \,  \, 
K \big( \U (w), \breve{\U} (w)  \big) \, \Big]  ~.
\label{nact} 
\ee
This 5D supersymmetric sigma-model 
respects all the geometric features of
its 4D, $\cN=1$ predecessor \cite{Zumino}
\be
S[\F, \bar \F] =  \int {\rm d}^4 x  \, {\rm d}^4 \q \,K(\Phi,
\, {\bar \Phi} )  ~,
\label{kahler}
\ee
where $K(A, \bar A)$ is the  K\"ahler potential of some manifold $\cM$.
 The K\"ahler invariance of (\ref{kahler})
\be
K(\F, \bar \F) \quad \longrightarrow \quad K(\F, \bar \F) ~+~ 
\Big( \Lambda(\F)
\,+\,  {\bar \Lambda} (\bar \F) \Big)
\label{kahlerinv}
\ee
turns into 
\be
K(\U, \breve{\U})  \quad \longrightarrow \quad K(\U, \breve{\U}) ~+~
\Big(\Lambda(\U) \,+\, {\bar \Lambda} (\breve{\U} ) \Big)
\ee
for the model (\ref{nact}). 
A holomorphic reparametrization $A^I
\mapsto
f^I \big( A \big)$ of the K\"ahler manifold has the following
counterparts
\be
\F^I  \quad  \mapsto   \quad f^I \big( \F \big) ~, \qquad \qquad 
\U^I (w) \quad  \mapsto  \quad f^I \big (\U(w) \big)
\ee
in the 4D and 5D cases, respectively. Therefore, the physical
superfields of the 5D theory
\be
\U^I (w)\Big|_{w=0} ~=~ \F^I ~,\qquad  \quad
 \frac{ {\rm d} \U^I (w) }{ {\rm d} w} \Big|_{w=0} ~=~ \G^I ~,
\label{geo3} 
\ee
should be regarded, respectively, as a coordinate of the K\" ahler
manifold and a tangent vector at point $\F$ of the same manifold. 
That is why the variables $(\F^I, \G^J)$ parametrize the tangent 
bundle $T\cM$ of the K\"ahler manifold $\cM$.

The auxiliary superfields $\U_2, \U_3, \dots$, and their
conjugates,  can
be eliminated  with the aid of the 
corresponding algebraic equations of motion
\be
 \oint {{\rm d} w} \,w^{n-1} \, \frac{\pa K(\U, \breve{\U} 
) }{\pa \U^I} = 0~,
\qquad n \geq 2 ~.               
\label{int}
\ee
Their elimination  can be  carried out
using the ansatz \cite{Kuz}
\bea
\U^I_n = \sum_{p=o}^{\infty} 
G^I{}_{J_1 \dots J_{n+p} \, \bar{L}_1 \dots  \bar{L}_p} (\F, {\bar \F})\,
\G^{J_1} \dots \G^{J_{n+p}} \,
{\bar 
\G}^{ {\bar L}_1 } \dots {\bar \G}^{ {\bar L}_p }~, 
\qquad n\geq 2~.
\eea
Upon elimination of the auxiliary superfields,\footnote{As
explained in \cite{GK}, the auxiliary superfields can be eliminated
only perturbatively for general K\"ahler manifolds. 
This agrees with a theorem proved in \cite{cotangent} that, 
for a K\"ahler manifold $\cM$, a  canonical hyper-K\"ahler structure 
 exists, in general, on an open neighborhood of the zero section 
of the cotangent bundle $T^*\cM$. It was further demonstrated 
in \cite{GK} that the auxiliary superfields can be eliminated 
in the case of compact K\"ahler symmetric spaces.}
the action 
(\ref{nact}) takes the form
\bea
S_{{\rm tb}}[\F, \bar \F, \G, \bar \G]  
&=& \int {\rm d}^5 x \, {\rm d}^4 \q \, \Big\{\,
K \big( \F, \bar{\F} \big) - g_{I \bar{J}} \big( \F, \bar{\F} 
\big) \G^I {\bar \G}^{\bar{J}} 
\non\\
&&\qquad +
\sum_{p=2}^{\infty} \cR_{I_1 \cdots I_p {\bar J}_1 \cdots {\bar 
J}_p }  \big( \F, \bar{\F} \big) \G^{I_1} \dots \G^{I_p} {\bar 
\G}^{ {\bar J}_1 } \dots {\bar \G}^{ {\bar J}_p }~\Big\}~, 
\eea
where the tensors $\cR_{I_1 \cdots I_p {\bar J}_1 \cdots {\bar 
J}_p }$ are functions of the Riemann curvature $R_{I {\bar 
J} K {\bar L}} \big( \F, \bar{\F} \big) $ and its covariant 
derivatives.  Each term in the action contains equal powers
of $\G$ and $\bar \G$, since the original model (\ref{nact}) 
is invariant under rigid U(1)  transformations
\be
\U(w) ~~ \mapsto ~~ \U({\rm e}^{{\rm i} \a} w) 
\quad \Longleftrightarrow \quad 
\U_n(z) ~~ \mapsto ~~ {\rm e}^{{\rm i} n \a} \U_n(z) ~.
\label{rfiber}
\ee

${}$For the theory with action  $S_{{\rm tb}}[\F, \bar \F, \G, \bar \G]  $, 
we can develop a dual formulation involving only chiral superfields 
and their  conjugates as the dynamical variables. 
Consider the first-order action 
\bea
&&S_{{\rm tb}}[\F, \bar \F, \G, \bar \G] 
- \int {\rm d}^5 x \,\Big\{ 
\int {\rm d}^2 \q\, \J_I \Big( \pa_5\,  \F^I +
{1\over 4} {\bar D}^2 \, \G^I \Big)
~+~{\rm c.c.} \Big\} \non \\
&=& S_{{\rm tb}}[\F, \bar \F, \G, \bar \G] 
+ \int {\rm d}^5 x \,\Big\{ 
\int {\rm d}^4 \q \, \J_I \,\G^I
-\int {\rm d}^2 \q\, \J_I \pa_5\,  \F^I 
~+~{\rm c.c.} \Big\} ~,
\eea
where the tangent vector $\G^I$ is now  complex unconstrained, 
while the one-form $\J_I$ is chiral, ${\bar D}_{\dt \a} \J_I =0$.
Upon elimination of $\G$ and $\bar \G$, with the aid of their equations of 
motion, the action turns into $S_{{\rm cb}}[\F, \bar \F, \J, \bar \J]$.
Its target space is  the cotangent 
bundle $T^*\cM$ of the K\"ahler manifold $\cM$. 

It is instructive to consider a free hypermultiplet  described
by the K\"ahler potential $K_{\rm free}(A, \bar A) ={\bar A}\,A$.  Then
\bea
S_{\rm free}[\U, \breve{\U}]  =  
\int {\rm d}^5 x \, {\rm d}^4 \q 
\sum_{n=0}^{\infty} (-1)^n {\bar \U}_n (z) \U_n
=\int {\rm d}^5 x \, {\rm d}^4 \q \,\Big( {\bar \F}\, \F 
-{\bar \G}\, \G \Big)~+ ~\dots
\label{eqn:CMNfree}
\eea
Here the dots stand for the auxiliary superfields' contribution.
Now, eliminating the auxiliary superfields and dualizing 
$\G$ into a chiral scalar, one obtains the action
 for the free Fayet-Sohnius 
hypermultiplet, equation (\ref{FS-actionN=1}).

\sect{Vector Multiplet in Projective Superspace}
\label{section:seven}

In the Abelian case, the gauge
transformation (\ref{lambda2}) simplifies 
\be
\d \cV^{++} = - D^{++} \lambda~, \qquad 
D^+_{\hat \a} \lambda=0~, \qquad 
\breve{\lambda}=\lambda~.
\label{lambda3}
\ee
The field strength (\ref{W2})  
also simplifies
\be
\cW = \frac{\rm i}{8} \int {\rm d}u \, 
 ({\hat D}^-)^2 \, \cV^{++}~. 
\label{f-strength}
\ee
It is easy to see that $\cW$ is gauge invariant.  

The gauge freedom (\ref{lambda3}) can be used to 
choose the supersymmetric Lorentz gauge \cite{GIKOS}
\be
D^{++} \cV^{++} =0~.
\label{Lorentz}
\ee 
In other words, in this gauge  
$\cV^{++}$ becomes a real O(2) multiplet, 
\be 
\cV^{++}={\rm i}\, u^{+\1}u^{+\2}\, V(z,w)~, 
\qquad 
V(z,w)=  {1\over w}\, \varphi (z)+ V(z)-w \,
\bar \varphi (z)~.
\label{Lorentz2}
\ee
Since $\cW$ is gauge invariant, 
for its  evaluation one can use 
any  potential $\cV^{++}$  from the same gauge orbit, 
in particular the  one obeying  the gauge condition (\ref{Lorentz}).
This Lorentz gauge is particularly useful for our consideration. 
Using the relation  (\ref{Diamond}) and noting that
$|u^{+\1}|^2 = (1+w {\bar w})^{-1}$, we can rewrite 
$\cW$ in the form
\be
\cW = \hf \int {\rm d}u \, 
\cP(w)  \,V(z,w)~. 
\label{f-strength2}
\ee
This can be further transformed to 
\be
\cW = {1\over 4\pi {\rm i}} \oint {{\rm d} w\over w} \,
\cP(w)  \, V(z,w) ~.
\label{f-strength3}
\ee
Indeed, the consideration in Appendix C 
justifies the following identity
\bea
\lim_{R\to \infty} \lim_{\e\to 0} \int {\rm d} u \,\f_{R,\e}(u)
= {1\over 2\pi {\rm i}} \oint {{\rm d} w\over w} \, \f(w)~,
\label{easy1}
\eea
with the regularization  $\f_{R,\e}(u)= \f_{R,\e}(w,\bar w)$  
of a function $\f(w) $ holomorphic on ${\mathbb C}^{*}$
defined according to (\ref{regularization}).
Since the integrand on the right of (\ref{f-strength2}) 
is, by construction, 
a smooth scalar field on $S^2$, we obvoiusly have 
\be
\int {\rm d}u \,  \cP(w)  \,V(z,w) 
= \lim_{R\to \infty} \lim_{\e\to 0} \int {\rm d} u \,
\cP(w)  \,V_{R,\e}(z,u)~.
\label{easy2}
\ee

The representation (\ref{f-strength3}) allows one to obtain 
a new formulation for the vector multiplet. 
Let $\Lambda(z,w)$ be an arctic  multiplet
\be
\Lambda (z,w) = \sum_{n=0}^{\infty} \Lambda_n (z) \,w^n~,
\qquad 
\nabla_{\hat \a} (w) \Lambda(z, w) = 0~,
\ee
and $\breve{\Lambda} (z,w)$ its smile-conjugate. 
It then immediately follows that
\be 
 \oint {{\rm d} w\over w} \, \cP(w)  \,\Lambda (z,w)
= \oint {{\rm d} w\over w} \,\cP(w)  \breve{\Lambda} (z,w) =0~.
\ee
Now, introduce a real tropical multiplet $V(z,w)$, 
\be 
V (z, w) = \sum_{n=-\infty}^{+\infty} 
V_n (z) \,w^n~, 
\qquad \nabla_{\hat \a} (w) V(z,w) = 0~,
\qquad
\bar{V}_n  =  
(-1)^n \,
V_{-n}~,
\label{tropical}
\ee
possessing  the gauge freedom
\be
\d V(z,w) = {\rm i}\Big( \breve{\Lambda} (z,w)-\Lambda (z,w) \Big)~.
\label{lambda4}
\ee
With such gauge transformations, 
eq. (\ref{f-strength3}) defines a gauge invariant 
field strength. Next, in accordance with the superfield structure 
of the tropical and arctic multiplets, the gauge freedom 
can be used to turn  $V(z,w)$ into a real  O(2) multiplet,  
i.e. to bring $V(z,w)$ to the form (\ref{Lorentz2}).  
We thus arrive at the projective superspace 
formulation\footnote{An alternative procedure to deduce
the  projective superspace formulation
for the 4D, $\cN=2$ vector multiplet 
from the corresponding harmonic superspace 
formulation can be found in 
 \cite{Kuzenko:1998xm}.}
for the vector multiplet \cite{projective}.

Now, we are in a position to evaluate  
the $\cN=1$ field strengths 
(\ref{N=1field-strengths})
in terms of the prepotentials $V_n$.
It follows from (\ref{Diamond}) that
\bea
\cF= { \cW}|
={1\over 4\pi {\rm i}} \oint {{\rm d} w\over w} 
\cP(w)\, V(w) {\Big|}
={1\over 2} \,\Big(\F+\bar \F  + \pa_5 V\Big)~,
\eea
where we have defined 
\be
\F=\frac 14\, \bar D^2 V_1|~, 
\qquad 
V =V_0|~.
\label{PhiV}
\ee
The spinor  field strength $\cW_\a$ is given 
by
\bea 
\cW_\a(z)= {D}^{\underline2}_\a
{\cal W} |
={1\over 4\pi {\rm i}} \oint {{\rm d} w\over w}\left( 
[{D}^{\underline2}_\a\,, \cP(w)]
+\cP(w) {D}^{\underline2}_\a
\right) V(w) {\Big|}~.
\eea
However, as 
$[{D}^{\underline2}_\a\,, \cP(w)]= 
w\, \pa_5 D_\a^{\1}$ 
and given that for any projective superfield $ \f(w)$
we have 
${D}^{\underline2}_\a \f(w)=w D_\a^{\1} \f(w)$,
this expression reduces to
\bea
\cW_\a
={1\over 4\pi {\rm i}} \oint {{\rm d} w\over w} \left(
{1\over 4} \bar D^2 D_\a \right) V(w)
{\Big|} ={1\over 8} \bar D^2 D_\a V~.
\eea
It can be seen that the gauge transformation 
(\ref{lambda4}) acts on the superfields in (\ref{PhiV}) 
as follows:
\be
\label{eqn:AbelianGaugeTrans}
\d V = {\rm i} \,( {\bar \Lambda}-\Lambda)~, \qquad
\d \F = {\rm i} \,\pa_5 \, \Lambda \qquad \quad 
\Lambda =\Lambda_1|~.
\ee

The approach presented in this section 
can be applied to reformulate
the supersymmetric Chern-Simons theory
(\ref{CS2}) in projective superspace, 
and the possibility for this 
is based on the following observation.
Let  $\cL^{++} $ be a linear multiplet, that is 
a real analytic superfield obeying the constraint
$D^{++} \,\cL^{++} =0$. 
Then, the functional 
$$
\int {\rm d} \z^{(-4)} \, \cV^{++} \,\cL^{++} 
$$
is invariant under the gauge transformations
(\ref{lambda3}). We can further represent 
$\cL^{++}=({\rm i} u^{+\1}u^{+\2}) L(z,w)$, 
where $\cL(z,w)$ is a real O(2) multiplet.
Then the functional
$$
-\frac{1}{2\pi \,{\rm i} } 
 \int {\rm d}^5 x \, {\rm d}^4 \q \, 
\oint \frac{{\rm d}w}{ w}\, V(w)\,L(w)\Big|
$$
is invariant under the gauge transformations
(\ref{lambda4}).

In the case of Chern-Simons theory
(\ref{CS2}), the role of $\cL^{++} $ is played by 
the gauge-invariant superfield $(12g^2)^{-1} \,\cG^{++}$,  
with $\cG^{++}$ defined in (\ref{YML}).  
With the real O(2) multiplet $G(z,w)$  
introduced by 
\be
\cG^{++}= ({\rm i} u^{+\1}u^{+\2}) \, G(z,w)~,
\qquad G(w) = -{1\over w} \, \J+K+ w\,  \bar \J~,
\label{sYMRed}
\ee 
the Chern-Simons theory (\ref{CS2}) is equivalently 
described by the action 
\be
12g^2\, S_{\rm CS} = -
\frac{1}{2\pi \,{\rm i} } 
 \int {\rm d}^5 x \, {\rm d}^4 \q \, 
\oint \frac{{\rm d}w}{ w}\, V(w)\,G(w)\Big|
\equiv 12g^2  \int {\rm d}^5 x \, {\bm \cL}_{\rm CS}
~.
\ee
Direct evaluation of $\J$ and $K$ gives
\bea
\J &=&-\cW^\a \cW_\a+{1\over 2} \bar D^2 (\cF^2)~, \non \\
K&=&-{\cal F}D^\a {\cal W}_\a-2(D^\a {\cal F}){\cal W}_\a 
+{\rm c.c.}+2\pa_5({\cal F}^2)~. 
\label{eqn:sYMRed2}
\eea
These results lead to 
\bea
12g^2\, {\bm \cL}_{\rm CS}
=\int {\rm d}^2 \q \, \F \, {\cal W}^\a{\cal W}_\a
&+&\int {\rm d}^4 \q \, V\, 
\left[ {\cal F} D^\a {\cal W}_\a +2(D^\a {\cal F}) {\cal W}_\a\right]
~+~{\rm c.c.} \non \\
&+& 4 \int {\rm d}^4 \q \, {\cal F}^3~.
\label{eqn:CSaction}
\eea
Here we have chosen to present the answer in the form 
${\rm( potential)}\times{\rm(fieldstrength)}\times {\rm(fieldstength)}$ 
analogously to the standard representation of the bosonic Chern-Simons
action.\footnote{The 
result presented here was given previously in the first reference 
of  \cite{hybrid}. In comparing the results one should keep in mind that 
terms such as $\int {\rm d}^4 \theta \, V 
\left[ \frac 12 (\Phi +\bar \Phi)D^\alpha {\cal W}_\alpha
+D^\alpha\Phi {\cal W}_\alpha \right]+{\rm c.c.}$ 
can be rewritten to look like 
$\int {\rm d}^2 \q\, \Phi {\cal W}^\alpha {\cal W}_\alpha +{\rm c.c.}$, 
thereby changing the appearance of the action.}

The structure of the  superspace action obtained
is the following. 
The first and second line of (\ref{eqn:CSaction}) are separately invariant 
under the gauge transformation (\ref{eqn:AbelianGaugeTrans}) 
up to surface terms as is easily seen. The relative factor of 4 is fixed 
by five-dimensional Lorentz invariance. This could be derived either from 
the component projection or, less painfully, by checking the five-dimensional 
mass-shell condition on the super-fieldstrengths using their equations of motion
together with their Bianchi identities. Finally, 
under the shift $\Phi\mapsto \Phi+1$,
the action shifts by 
$S_{\rm CS}\mapsto S_{\rm CS}+S_{\rm YM}+{\rm surface~ term,}$ 
where $S_{\rm YM}$ is the 5D Yang-Mills action (\ref{eqn:SimpleYM}) 
with the proper normalization.

${}$For completeness, we also present here projective superspace extensions
of the vector multiplet mass term and the Fayet-Iliopoulos term 
(their harmonic superspace form is given in  \cite{GIKOS}).
The vector multiplet mass term is
\be
-m^2 \int {\rm d} \z^{(-4)} \, (\cV^{++} )^2 \quad 
\longrightarrow \quad
\frac{m^2}{2\pi \,{\rm i} } 
 \int {\rm d}^5 x \, {\rm d}^4 \q  
\oint \frac{{\rm d}w}{ w}\, V^2(w)\Big|~.
\ee
The gauge invariant Fayet-Iliopoulos term is
\be
\int {\rm d} \z^{(-4)} \, c^{++}\, \cV^{++} 
 \quad 
\longrightarrow \quad
-\frac{1}{2\pi \,{\rm i} } 
 \int {\rm d}^5 x \, {\rm d}^4 \q  
\oint \frac{{\rm d}w}{ w}\, c(w)V(w)\Big|~,
\ee
where  $c^{++} = c^{ij} u^+_i u^+_j$, with a constant real iso-vector $c^{ij}$. Defining 
$c^{++}={\rm i} u^{+\underline 1}u^{+\underline 2} \,c(w)$, 
 with $c(w)=w^{-1} \bar \x_{\mathbb C} +\x_{\mathbb R}
-w  \x_{\mathbb C}$,  the FI action then reduces to 
\bea
\x_{\mathbb R} \int {\rm d}^5 x \, {\rm d}^4 \q \,V
+2{\rm Re} \,\Big(
\x_{\mathbb C} \int {\rm d}^5 x \, {\rm d}^2 \q \,\F \Big)~.
\eea

So far the considerations in this section have been restricted to the 
Abelian case. It is necessary to mention that the projective superspace approach \cite{projective} can be generalized 
to provide an elegant description 
of 5D super Yang-Mills theories, which is very similar 
to the well-known description of 4D, $\cN=1$ supersymmetric 
theories. In particular, the Yang-Mills supermultiplet is 
described by a real Lie-algebra-valued tropical superfield
$V(z,w)$ with the gauge transformation
\be 
{\rm e}^{V(w)} ~\to ~ {\rm e}^{{\rm i} \breve{\Lambda}(w)} \,
{\rm e}^{V(w)} \,{\rm e}^{-{\rm i} \Lambda(w)} ~,
\ee
which is the non-linear generalization of 
the Abelian gauge transformation (\ref{lambda4}).
The hypermultiplet sector  is described by an arctic superfield
$\U(z,w)$ and its conjugate, with the gauge
transformation 
 \be 
\U(w)~\to ~{\rm e}^{{\rm i} \Lambda(w)} \,\U(w)~.
\ee
The hypermultiplet gauge--invariant action is
\be
S[\U, \breve{\U}, V ]  = \frac{1}{2\pi {\rm i}} \, \oint \frac{{\rm d}w}{w} 
 \int {\rm d}^5 x \, {\rm d}^4 \q \,  
 \breve{\U} (w)\, {\rm e}^{V(w)} \, \U(w)  \ ~.
\ee

\sect{Conclusion}
\label{section:eight}

In the present  paper we have developed the manifestly 
supersymmetric approach to five-dimensional globally 
supersymmetric gauge theories. 
It is quite satisfying that 5D superspace techniques 
provide a universal setting to formulate all such theories 
in a compact, transparent and elegant form, similarly 
to the four-dimensional $\cN=1$ and $\cN=2$ theories.
We believe that these techniques are not only elegant
but, more importantly,  are useful.
In particular, these techniques may be useful for 
model building in the context of supersymmetric 
brane-world scenarios. The two examples of 
supersymmetric nonlinear sigma-models,
which were 
constructed in section 6, clearly demonstrate 
the power of the 5D superspace  approach. 

Five-dimensional super Yang-Mills theories
possess interesting properties at the quantum 
level \cite{Seiberg}. Further insight into their 
quantum mechanical structure may be obtained
by carrying out explicit supergraph calculations.
Supersymmetric Chern-Simons theories (\ref{CS})
are truly  interesting  in this respect.

{\bf Note Added:} After this paper was posted to the
hep-th archive, we were informed of a related 
interesting work on 6D, $\cN=(1,0)$ supersymmetric
field theories  
\cite{GMP}. 
\vskip.5cm

\noindent
{\bf Acknowledgements:}\\
SMK is grateful to Jim Gates and the Center for String and 
Particle Theory at the  University of Maryland, where this project 
was conceived, for hospitality.  The work of SMK is supported in part 
by the Australian Research Council. The work of WDL is supported 
by the University of Maryland Center for String and Particle Theory.

\begin{appendix}
\sect{5D Notation and Conventions}
\label{section:A}
Our 5D notation and conventions are very similar 
to those  introduced in \cite{BB}.

The 5D gamma-matrices $\G_{\hat m} = ( \G_m, \G_5 )$, 
with $m=0,1,2,3$,
defined by 
\be
\{ \G_{\hat m} \, , \,\G_{\hat n} \} 
= - 2 \eta_{\hat m \hat n} \,{\bf 1}~, \qquad
(\G_{\hat m} )^\dagger = \G_0 \, \G_{\hat m} \, \G_0 
\ee
are chosen in accordance with \cite{WB,BK}
\bea
(\G_m ){}_{\hat \a}{}^{\hat \b}=
\left(
\begin{array}{cc}
0 ~ &  (\s_m)_{\a \dt \b} \\
(\tilde{\s}_m)^{\dt \a \b} ~ & 0  
\end{array}
\right)~, \qquad
(\G_5 ){}_{\hat \a}{}^{\hat \b}=
\left(
\begin{array}{cc}
-{\rm i} \,\d_\a{}^\b~ &  0 \\
0 ~ & {\rm i}\, \d^{\dt \a}{}_{\dt \b}
\end{array}
\right)~,
\eea
such that $\G_0 \G_1 \G_2 \G_3 \G_5 ={\bf 1}$. 
The charge conjugation matrix, $C = (\ve^{\hat \a \hat \b})$, 
and its inverse, $C^{-1} = C^\dag =(\ve_{\hat \a \hat \b})$ 
are defined by 
\bea
C\,\G_{\hat m} \,C^{-1} = (\G_{\hat m}){}^{\rm T}~,
\qquad 
\ve^{\hat \a \hat \b}=
\left(
\begin{array}{cc}
 \ve^{\a \b} &0 \\
0& -\ve_{\dt \a \dt \b}    
\end{array}
\right)~, \quad
\ve_{\hat \a \hat \b}=
\left(
\begin{array}{cc}
 \ve_{\a \b} &0 \\
0& -\ve^{\dt \a \dt \b}    
\end{array}
\right)~.
\eea
The antisymmetric matrices $\ve^{\hat \a \hat \b}$ and
$\ve_{\hat \a \hat \b}$ are used to raise and lower the four-component 
spinor indices.

A Dirac spinor, $\J=(\J_{\hat \a}) $, and its Dirac conjugate, 
$\J =({\bar \J}^{\hat \a}) = \J^\dag \,\G_0$, look like
\bea
\J_{\hat \a} =   
\left(
\begin{array}{c}
\j_\a \\
{\bar \f}^{\dt \a}    
\end{array}
\right)~, \qquad 
{\bar \J}^{\hat \a}= (\f^\a \,, \,{\bar \j}_{\dt \a})~.
\eea
One can now combine ${\bar \J}^{\hat \a}= (\f^\a , {\bar \j}_{\dt \a})$ and 
$\J^{\hat \a} = \ve^{\hat \a \hat \b} \J_{\hat \b} =(\j^\a , - {\bar \f}_{\dt \a} )$ 
into a SU(2) doublet, 
\be
\J^{\hat \a}_i = (\J^\a_i,  -{\bar \J}_{\dt \a i} ) ~, \qquad
(\J^\a_i)^* = {\bar \J}^{\dt \a i}~, \qquad 
i = \1 , \2 ~,   
\ee 
with $\J^\a_{\1} = \f^\a $ and $\J^\a_{\2} = \j^\a $.
It is understood that the SU(2) indices are raised and lowered 
by $\ve^{ij} $ and  $\ve_{ij} $, $\ve^{\1 \2} =  \ve_{\2 \1} =1$, 
in the standard fashion: $\J^{\hat \a i} = \ve^{ij} \J^{\hat \a}_j$.
The  Dirac  spinor $\J^i = ( \J^i_{\hat \a}  )$
satisfies the pseudo-Majorana condition
${\bar \J}_i{}^{\rm T} = C \,   \J_i$.
This will be concisely represented as
\be
(\J^i_{\hat \a} )^* = \J^{\hat \a}_i~.
\ee

With the definition $\S_{\hat m \hat n} 
=-\S_{\hat n \hat m} = -{1 \over 4}
[\G_{\hat m} , \G_{\hat n} ] $, the matrices 
$\{ {\bf 1}, \G_{\hat m} , \S_{\hat m \hat n} \} $
form a basis in the space of  $4 \times 4$ matrices. 
The matrices $\ve_{\hat \a \hat \b}$ and 
$(\G_{\hat m})_{\hat \a \hat \b}$ are antisymmetric, 
$\ve^{\hat \a \hat \b}\, (\G_{\hat m})_{\hat \a \hat \b} =0$, 
while the matrices $(\S_{\hat m \hat n})_{\hat \a \hat \b}$ 
are symmetric.  
Given a 5-vector $V^{\hat m}$ and an 
antisymmetric tensor $F^{\hat m \hat n} = -F^{\hat n \hat m}$,
we can equivalently represent  them as the 
bi-spinors $V = V^{\hat m} \,  \G_{\hat m}$
and $F = \hf F^{\hat m \hat n}\, \S_{\hat m \hat n} $
with the following symmetry properties
\bea 
V_{\hat \a \hat \b} &=& -V_{\hat \b \hat \a} ~, 
\quad \ve^{\hat \a \hat \b}\, V_{\hat \a \hat \b} =0~, 
 \qquad \quad
F_{\hat \a \hat \b} = F_{\hat \b \hat \a} ~. 
\eea
The two equivalent descriptions 
$ V_{\hat m} \leftrightarrow V_{\hat \a \hat \b}$ and 
and $ F_{\hat m \hat n} \leftrightarrow F_{\hat \a \hat \b}$
are explicitly described as follows:
\bea 
V_{\hat \a \hat \b} = V^{\hat m} \, ( \G_{\hat m})_{\hat \a \hat \b}~,
\quad && \quad 
V_{\hat m}  = -{1 \over 4} \,( \G_{\hat m})^{\hat \a \hat \b}\,
V_{\hat \a \hat \b}~, \non \\
F_{\hat \a \hat \b} = \hf F^{\hat m \hat n} 
(\S_{\hat m \hat n})_{\hat \a \hat \b}~, \quad && \quad 
F_{\hat m \hat n}  = (\S_{\hat m \hat n})^{\hat \a \hat \b} \,
F_{\hat \a \hat \b} ~.
\eea
These results can be easily checked using the identities
(see e.g. \cite{Z}):
\bea
 \ve _{\hat \a \hat \b \hat \g \hat \d} 
&=& \ve_{\hat \a \hat \b} \, \ve_{\hat \g \hat \d}
+ \ve_{\hat \a \hat \g} \, \ve_{\hat \d \hat \b}
+\ve_{\hat \a \hat \d} \, \ve_{\hat \b \hat \g}~,  \non \\
 \ve_{\hat \a \hat \g} \, \ve_{\hat \b \hat \d}
-\ve_{\hat \a \hat \d} \, \ve_{\hat \b \hat \g}
&=&-\hf \,  ( \G^{\hat m})_{\hat \a \hat \b}\,
( \G_{\hat m})_{\hat \g \hat \d}
+\hf \, \ve _{\hat \a \hat \b}\, \ve_{ \hat \g \hat \d} ~,
\eea
and therefore 
\be
 \ve _{\hat \a \hat \b \hat \g \hat \d} 
=\hf \,( \G^{\hat m})_{\hat \a \hat \b}\,
( \G_{\hat m})_{\hat \g \hat \d}
+\hf \, \ve _{\hat \a \hat \b} \, \ve_{ \hat \g \hat \d} ~,
\ee
with 
$ \ve _{\hat \a \hat \b \hat \g \hat \d} $ the completely 
antisymmetric fourth-rank tensor.

Complex conjugation gives 
\be 
(\ve_{\hat \a \hat \b})^* = - \ve^{\hat \a \hat \b}~,
\qquad 
(V_{\hat \a \hat \b})^* = V^{\hat \a \hat \b}~,
\qquad 
(F_{\hat \a \hat \b})^* = F^{\hat \a \hat \b}~,
\ee
provided   $V^{\hat m}$ and  $F^{\hat m \hat n} $ are real.

The conventional 5D simple superspace ${\mathbb R}^{5|8}$ 
is parametrized  
by  coordinates  $ z^{\hat A} = (x^{\hat a},  \q^{\hat \a}_i )$. 
Then,  a hypersurface $x^5 ={\rm const}$ in ${\mathbb R}^{5|8}$ 
can be identified  with the 4D, $\cN=2$  superspace 
${\mathbb R}^{4|8}$ parametrized by 
\be
z^{A} = (x^a,  \q^\a_i , {\bar \q}_{\dt \a}^i)~, \qquad 
(\q^\a_i )^* = {\bar \q}^{\dt \a i}~.
\ee
The Grassmann coordinates of ${\mathbb R}^{5|8}$ and 
${\mathbb R}^{4|8}$
are related to each other as follows:
\bea
\q^{\hat \a}_i = ( \q^\a_i , - {\bar \q}_{\dt \a i})~, 
\qquad
\q_{\hat \a}^i =   
\left(
\begin{array}{c}
\q_\a^i \\
{\bar \q}^{\dt \a i}    
\end{array}
\right)~.
\eea
Interpreting $x^5$ as a central charge variable, 
one can view ${\mathbb R}^{5|8}$ as a 4D, $\cN=2$ 
central charge superspace (see below).

The flat covariant derivatives 
$D_{\hat A} = ( \pa_{\hat a} , D^i_{\hat \a}) $ obey the
algebra
\be
\{D^i_{\hat \a} \, , \, D^j_{\hat \b} \} = -2{\rm i} \,
\ve^{ij}\,
\Big( (\G^{\hat c} ){}_{\hat \a \hat \b} \, \pa_{\hat c} 
+ \ve_{\hat \a \hat \b} \,\D \Big)~,
\qquad [ D^i_{\hat \a} \, , \, \pa_{\hat b} ] = 
[ D^i_{\hat \a} \, , \, \D ] =0~, 
\ee
or equivalently 
\be 
[ D_{\hat A} \, ,\, D_{\hat B} \} = T_{\hat A \hat B}{}^{\hat C} \, 
D_{\hat C}  + C_{\hat A \hat B} \, \D~,
\label{flat}
\ee
with $\D$ the central charge.
The spinor covariant derivatives are 
\be
D^i_{\hat \a} = \frac{\pa}{\pa \q^{\hat \a}_i} 
- {\rm i} \, (\G^{\hat b} ){}_{\hat \a \hat \b} \, \q^{\hat \b i}
\, \pa_{\hat b}
- {\rm i} \, \q_{\hat \a}^i \, \D~.
\ee
One can relate the operators
\bea
D^i  \equiv (D^i_{\hat \a} )
= \left(
\begin{array}{c}
D_\a^i \\
{\bar D}^{\dt \a i}    
\end{array}
\right)~, 
\qquad 
{\bar D}_i \equiv (D^{\hat \a}_i ) = 
(D^\a_i \,, \, -{\bar D}_{\dt \a i}) 
\label{con}
\eea 
to the 4D, $\cN=2$ covariant derivatives 
$D_A = (\pa_a , D^i_\a , {\bar D}^{\dt \a}_i )$
 where \cite{GIKOS,Ohta2}
\bea
 D^i_\a &=& \frac{\pa}{\pa \q^{\a}_i}
+ {\rm i} \,(\s^b )_{\a \bd} \, {\bar \q}^{\dt \b i}\, \pa_b
-{\rm i}  \,\q^i_\a \,(\D + {\rm i} \, \pa_5) ~, \non \\
{\bar D}_{\dt \a i} &=& 
- \frac{\pa}{\pa {\bar \q}^{\dt \a i}} 
- {\rm i} \, \q^\b _i (\s^b )_{\b \dt \a} \,\pa_b
-{\rm i}  \,{\bar \q}_{\dt \a i} \,(\D - {\rm i} \, \pa_5) ~. 
\label{4D-N2covder1}
\eea
These operators obey the anti-commutation relations
\bea
\{D^i_{\a} \, , \, D^j_{ \b} \} &=& -2{\rm i} \,
\ve^{ij}\, \ve_{\a \b} \,(\D + {\rm i} \, \pa_5) ~,
\qquad 
\{{\bar D}_{\dt \a i} \, , \, {\bar D}_{\dt  \b j} \} = 2{\rm i} \,
\ve_{ij}\, \ve_{\dt \a \dt \b} \,(\D - {\rm i} \, \pa_5) ~, \non \\
\{D^i_{\a} \, , \, \bar D_{ \dt \b j} \} &=& -2{\rm i} \, \d^i_j\,
(\s^c )_{\a \dt \b} \,\pa_c ~,
\label{4D-N2covder2}
\eea
which correspond to  the 4D, $\cN=2$ supersymmetry algebra with 
a complex central charge (see also  \cite{DIKST}).

In terms of the operators (\ref{con}), the operation 
of complex conjugation acts as follows
\be
(D^i \, F)^\dag \, \G_0 = -(-1)^{\e (F)} \, 
{\bar D}_i \, F^*~, 
\ee
with $F$ an arbitrary superfield and $\e (F)$ 
its Grassmann parity.
This  can be concisely represented as 
\be 
(D^i_{\hat \a} \, F)^* =  -(-1)^{\e (F)} \, 
D_i^{\hat \a} \, F^*~.
\ee

\sect{Tensor Fields on the Two-Sphere}
\label{section:B}
In this appendix we recall, following \cite{Kuzenko:1998xm},
the well-known one-to-one
correspondence between smooth tensor fields on 
$ S ^2 = {\rm SU}(2)/{\rm U}(1)$ and
smooth scalar functions over SU(2) with definite U(1) charges.
The two-sphere is obtained from SU(2) by factorization
with respect to the equivalence relation
\be
u^{+i} \sim {\rm e}^{{\rm i} \varphi} u^{+i}  \qquad \varphi \in
{\mathbb R} \;.
\label{er}
\ee

We start by introducing two open charts 
forming an atlas on SU(2) which, 
upon identificationon (\ref{er}), leads to 
a useful atlas on $ S ^2$. The north patch is defined by
\be
 u^{+\1} \neq 0~,
\label{np}
\ee
and here we can represent
\bea
u^{+i} = u^{+\1}\, w^i ~,& \qquad & 
w^i = (1, u^{+\2} / u^{+\1}) = (1,w)~,
 \non \\
u^-_i = \overline{u^{+\1}} \,{\bar w}_i~,
 & \qquad & {\bar w}_i = (1, {\bar w})~,
\qquad |u^{+\1}|^2 = (1+ w {\bar w})^{-1}~.
\eea
The south patch is defined by
\be
 u^{+\2} \neq 0~,
\label{sp}
\ee
and here we have
\bea
u^{+i} = u^{+\2} \,y^i~, & \qquad & y^i 
= (u^{+\1} / u^{+\2},1) = (y,1)~,
 \non \\
u^-_i = \overline{u^{+\2}} \,{\bar y}_i ~,
& \qquad & {\bar y}_i = ({\bar y},1)~,
\qquad |u^{+\2}|^2 = (1+ y {\bar y})^{-1}~.
\eea
In the overlap of the two charts we have
\be
u^{+i} = \frac{ {\rm e}^{{\rm i} \a} }{ \sqrt{(1+ w {\bar w})} } \;w^i =
\frac{ {\rm e}^{{\rm i} \b} }{ \sqrt{(1+ y {\bar y})} } \;y^i~,
\ee
where
\be
y = \frac{1}{w} ~,\qquad \quad {\rm e}^{{\rm i} \b} =
\sqrt{\frac{w}{\bar w}} \;{\rm e}^{{\rm i} \a}~.
\ee
The variables $w$ and $y$ are seen to be local complex 
coordinates on $ S ^2$ considered as the Riemann sphere, 
$ S ^2 = {\mathbb C} \cup \{\infty\}$;
the north chart $U_{\rm N} = {\mathbb C}$ is parametrized by $w$
and the south patch $U_{\rm  S}  = {\mathbb C}\,{}^* \cup \{\infty\}$ is
parametrized by $y$.

Along with $w^i$ and ${\bar w}_i$, we often use their
counterparts with lower (upper) indices
\be
w_i = \ve_{ij} w^j = (-w, 1)~, 
\qquad {\bar w}^i = \ve^{ij} {\bar w}_j
= ({\bar w}, -1)~, \qquad \overline{w_i} = - {\bar w}^i~,
\ee
and similar for $y_i$ and ${\bar y}^i$.

Let $\J^{(p)}(u)$ be a smooth function on SU(2) 
with U(1)-charge $p$ chosen, 
for definiteness, to be non-negative, $p \geq 0$.
Such a function possesses a convergent Fourier series of the form
\be
\J^{(p)}(u) = \sum_{n=0}^{\infty} \J^{(i_1 \cdots i_{n+p} j_1 \cdots j_n)}
u^+_{i_1} \cdots u^+_{i_{n+p}} u^-_{j_1} \cdots u^-_{j_n} ~,
\qquad p \geq 0~.
\label{smoothfunction}
\ee
In the north patch we can write
\bea
\J^{(p)}(u) &=& (u^{+\1})^p \; \J^{(p)}_{{\rm N}}(w, {\bar w}) ~,\non \\
\J^{(p)}_{{\rm N}}(w, {\bar w}) &=&
\sum_{n=0}^{\infty} \J^{(i_1 \cdots i_{n+p} j_1 \cdots j_n)} \;
\frac{ w_{i_1} \cdots w_{i_{n+p}} {\bar w}_{j_1} \cdots {\bar w}_{j_n} }
{(1 + w {\bar w})^n }~.
\label{fn}
\eea
In the south patch we have
\bea
\J^{(p)}(u) &=& (u^{+\2})^p \; \J^{(p)}_{{\rm S}}(y, {\bar y}) ~,\non \\
\J^{(p)}_{{\rm S}}(y, {\bar y}) &=&
\sum_{n=0}^{\infty} \J^{(i_1 \cdots i_{n+p} j_1 \cdots j_n)} \;
\frac{ y_{i_1} \cdots y_{i_{n+p}} {\bar y}_{j_1} \cdots {\bar y}_{j_n} }
{(1 + y {\bar y})^n }~.
\eea
${}$Finally, in the overlap of the two charts 
$\J^{(p)}_{{\rm N}}$ and
$\J^{(p)}_{{\rm S}}$ are simply related to each other
\be
\J^{(p)}_{{\rm S}}(y, {\bar y}) = \frac{1}{w^p}
\; \J^{(p)}_{{\rm N}}(w, {\bar w})~,
\ee
If we redefine
$$
\hat{\J}^{(p)}_{{\rm N}}(w, {\bar w})
= {\rm e}^{ {\rm i} p \pi/4 } \;\J^{(p)}_{{\rm N}}(w, {\bar w})~,
\qquad  \check{\J}^{(p)}_{{\rm S}}(y, {\bar y})=
{\rm e}^{ -{\rm i} p \pi/4 }\; \J^{(p)}_{{\rm S}}(y, {\bar y})~,
$$
the above relation takes the form
\be
\check{\J}^{(p)}_{{\rm S}}(y, {\bar y})
= \left( \frac{\pa y}{\pa w} \right)^{p/2} \;
\hat{\J}^{(p)}_{{\rm N}}(w, {\bar w})
\ee
and thus defines a smooth tensor field on $ S^2$.

\sect{Projective Superspace Action}
\label{section:C}

In this appendix we briefly demonstrate, 
following  \cite{Kuzenko:1998xm}, 
how to derive the projective superspace action
(\ref{integral-projective}) from 
the harmonic superspace action (\ref{integral-analyt}).
More details can be found in  \cite{Kuzenko:1998xm}.

Consider an arbitrary projective superfield
$\f(z,w)$, eq. (\ref{holom3}), which is allowed to be 
 singular only at $w=0$ and $w=\infty$
(i.e.  $\f(z,w)$ is
 holomorphic on the doubly punctured  sphere
 $S^2\backslash \{N\cup S\}$). 
It is possible to promote $\f(z,w)$
to a smooth analytic superfield over $S^2$ 
by smearing (regularizing) its singularities with 
functions used in the construction of the partition of
unity in differential geometry.

\begin{figure}[ht] 
   \centering
   \includegraphics[width=6in]{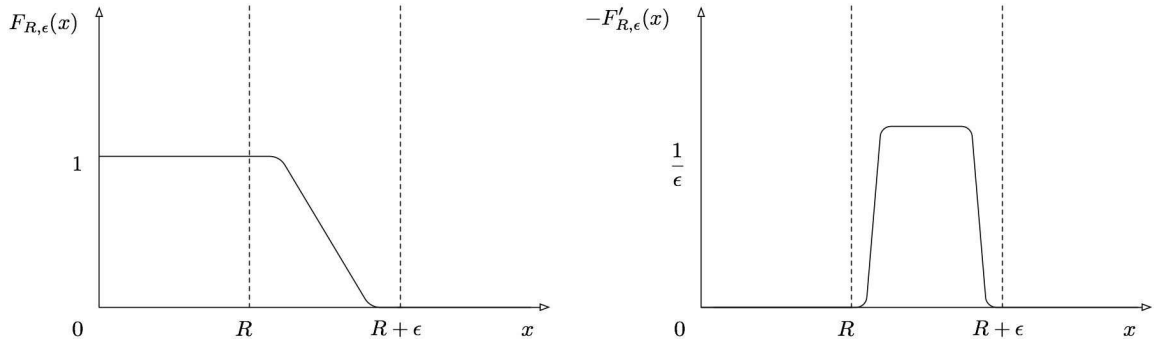}
   \caption{\small The function $F_{R,\e}(x)$ smoothy 
   interpolates from 1 to 0 in the region of width $\e$ starting at $R$. 
   It's derivative is a bump function with support $[R, R+\e]$ and unit area.}
   \label{fig:BumpFunction}
\end{figure}

Consider 
a smooth cut-off function $F_{R,\e}(x)$
sketched in figure \ref{fig:BumpFunction}.

This function extrapolates smoothly from unit magnitude 
to zero in a small region between $R$, with is assumed 
to be large number, and $R+\e$ where $\e$ is small. 
The derivative of this function localizes whatever 
it multiplies to this region 
and is normalized so that in the limit
\bea
\lim_{\e\to 0} F^\prime_{R,\e}(x)=-\d(x-R)
\eea
as a distribution.
Now, we can regularize 
the projective superfield
$\f(z,w) $
as follows:
\be
\f(z,w)  \quad \longrightarrow \quad   
\f_{R,\e} (z,w, \bar w)=
F_{R,\e} (|w|^{-1})\, \f(z,w) \, F_{R,\e}(|w|)~, 
\label{regularization}
\ee
and the result is a a smooth neutral 
analytic superfield
over the harmonic superspsace.
If $\f(z,w) $ is regular at $w=0$ or $w=\infty$, 
then the factor $F_{R,\e} (|w|^{-1}) $
 or $ F_{R,\e}(|w|)$ on the right of 
(\ref{regularization}) can be removed.

The above procedure can  also be used to generate charged 
analytic superfields. For instance, if $\Lambda(z,w) $ is
a real  projective superfield, $\breve{\Lambda} =\Lambda$, 
then the following superfields
\bea
L^{++}_{R,\e}(z,u) &=& 
{\rm i} u^{+\1}u^{+\2}\, F_{R,\e} (|w|^{-1}) \, L(z,w)\,  F_{R,\e}(|w|)
\equiv {\rm i} u^{+\1}u^{+\2}\,L_{R,\e}(z,w,\bar w) ~,
\label{RegLag+2}\\
L^{(+4)}_{R,\e}(z,u) &=& 
(u^{+\1}u^{+\2})^2 F_{R,\e} (|w|^{-1}) \, L(z,w)\,  F_{R,\e}(|w|)
\equiv (u^{+\1}u^{+\2})^2 L_{R,\e}(z,w,\bar w) 
\label{RegLag+4}
\eea
are real analytic superfields of charge $+2$ and $+4$, 
respectively. 
One can use $L^{(+4)}_{R,\e}(z,u)$ in the role of Lagrangian
 in (\ref{integral-analyt}).
In the final stages we will remove 
the regulator by taking first $\e\to 0$ and then $R\to \infty$. 

As is seen from (\ref{integral-analyt}) and (\ref{D4+-}), 
the analytic action involves
a square of $(\hat D^-)^2$, and therefore
 we sould express the operators $(\hat D^-)^2$ 
in local coordinates. 
What actually we need here is 
this operator acting on analytic or projective  superfields
$\F$ such that 
 $\nabla_\a(w) \F = {\bar \nabla}^{\dt \a} (w) \F=0$, 
with operators $ \nabla_\a$ and ${\bar \nabla}^{\dt \a}(w)$
defined in (\ref{nabla}).
 The analyticity allows us 
 to move all $\cD^{\2}_\a$ and $\bar \cD_{\2}^{ \dt \a}$ derivatives 
onto $\F$ and 
 rewrite them in terms of $D_\a^{\1}$ and $\bar D^{ \dt \a}_{\1}$. 
When this is done, 
 we find in local coordinates for an analytic $\F$
\bea
(\hat D^-)^2 \F&=&-4 \,(\overline{u^{+\1}})^2 \,
{(1+\bar w w)^2\over w}\,
\cP(w) \F~,
\label{HalfMeasure}
\eea
where we have defined the projective differential operator
\bea
\cP(w) ={1\over 4w} \, 
(\bar D_{\1})^2 + \pa_5 -{w\over 4} \, (D^{\1})^2~.
\label{Diamond}
\eea
It is worth pointing out that eq. (\ref{HalfMeasure})
 also holds in 
the presence of a non-vanishing central charge $\D$.
Using the analyticity of $\F$ again, it is easy to 
show that 
\bea
(\hat D^-)^4\F &=&(\overline{u^{+\1}})^4 \,
{(1+\bar w w)^4\over w^2}D^4\F
~+~\mbox{total derivatives}~,
\eea
with the $D^4$ operator defined by (\ref{D4}).  
The latter operator determines the projective superspace 
measure, see eq. (\ref{integral-projective2}). Finally making 
use of  the identity
\be 
{\rm d}u = \frac{{\rm d}^2w}{\p (1+ w \bar w )^2}~,
\ee
one obtains (note $|u^{+\1}|^2 = (1+w {\bar w})^{-1}$)
\be
\int {\rm d} \z^{(-4)} \,L^{(+4)}_{R,\e}(z,u)
=\frac{1}{\pi}
\int  {\rm d}^5 x \int \frac{{\rm d}^2 w}{(1+w {\bar w})^2} \,
D^4L_{R,\e} (z,w, {\bar w})\double{ \Big|}~.
\ee
Representing here 
$$ 
\frac{1}{(1+w {\bar w})^2} 
=- \frac{1}{w} \,
\pa_{{\bar w}} \, \frac{1}{(1+w {\bar w})}  
$$
and integrating by parts, 
one can then show 
\be
\lim\limits_{R \to \infty } \,
\lim\limits_{\e \to 0} \, 
\int {\rm d} \z^{(-4)} \,L^{(+4)}_{R,\e}(z,u)
= {1\over 2\pi {\rm i}}\int {\rm d}^5x 
\oint {{\rm d} w\over w} \,D^4 L(z, w)
\double{\Big|}~.
\label{ActionFullRed}
\ee
This is exactly the projective action.

The formalism developed in this appendix can be applied 
to obtain a nice representation for the supersymmetric action 
(\ref{cc-action1}) which is equivalent to 
\bea
\label{Action2}
S&=&{{\rm i}\over 4} \int {\rm d}^5x
\int {\rm d}u \,(\hat D^-)^2 L^{++} \double{\Big|}~,
\qquad \quad
D^+_{\hat \a} L^{++}=0~, \quad
D^{++} L^{++}=0~.
\eea
Representing $L^{++}= {\rm i} u^{+\1}u^{+\2}\, L(z, w)$ 
and using eq.  (\ref{HalfMeasure}), we obtain
\bea
\label{Action3}
S&=& \int {\rm d}^5x
\int {\rm d}u \,\cP(w)\,L(z, w)
\double{\Big|}~.
\eea
${}$Finally, making use of (\ref{easy1}) gives
\bea
S&=&{1\over 8\pi {\rm i}}\int {\rm d}^5x \oint {{\rm d} w\over w} 
 \left[{1\over w}\bar D^2-w D^2\right]
L(z, w)\double{\Big|}~.
\label{RedAction2}
\eea
As an example of the usefulness of such a form, 
we can consider the super Yang-Mills Lagrangian (\ref{SYM-action1}). 
A trivial contour integration in (\ref{RedAction2})
then immediately reproduces 
the action for this theory in 
reduced superspace (\ref{eqn:SimpleYM}).

\end{appendix}

\small{

}

\end{document}